\documentclass[twocolumn]{autart}    

\usepackage{graphicx}          
\usepackage{epsfig}
\usepackage{dsfont}
\usepackage{amsmath}
\usepackage{amsfonts}
\usepackage{subfigure}
\usepackage{color}
\usepackage{hyperref}
\usepackage{bm}
\usepackage{bbm}
\usepackage{caption}
\usepackage{multirow}
\usepackage{amssymb}
\usepackage{algorithm,algpseudocode}
\usepackage{diagbox}
\usepackage{tikz}
\usepackage{makecell}
\usepackage[super]{nth}
\newcommand{\Cross}{$\mathbin{\tikz [x=1.2ex,y=1.2ex,line width=.2ex, red] \draw (0,0) -- (1,1) (0,1) -- (1,0);}$}%

\newcommand{\Checkmark}{$\color{green}\checkmark$}
\begin{document}

\begin{frontmatter}

\title{Functional Bayesian Filter\thanksref{footnoteinfo}} 

\thanks[footnoteinfo]{This work was supported, in part, by DARPA Contracts N66001-10-C-2008, N66001-15-1-4054, and the Lifelong Learning Machines program from DARPA/MTO grant FA9453-18-1-0039. This paper was not presented at any IFAC meeting. Corresponding author Kan Li. Tel. +1 (352) 392-2682. 
	Fax +1 (352) 392-0044.}

\author{Kan Li}\ead{likan@ufl.edu} \and
\author{Jos\'{e} C. Pr\'{i}ncipe}\ead{principe@cnel.ufl.edu}

\address{Computational NeuroEngineering Laboratory, University of Florida, Gainesville, FL 32611 USA}  


\begin{abstract}                          
We present a general nonlinear Bayesian filter for high-dimensional state estimation using the theory of reproducing kernel Hilbert space (RKHS). Applying kernel method and the representer theorem to perform linear quadratic estimation in a functional space, we derive a Bayesian recursive state estimator for a general nonlinear dynamical system in the original input space. Unlike existing nonlinear extensions of Kalman filter where the system dynamics are assumed known, the state-space representation for the Functional Bayesian Filter (FBF) is completely learned from measurement data in the form of an infinite impulse response (IIR) filter or recurrent network in the RKHS, with universal approximation property. Using positive definite kernel function satisfying Mercer's conditions to compute and evolve information quantities, the FBF exploits both the statistical and time-domain information about the signal, extracts higher-order moments, and preserves the properties of covariances without the ill effects due to conventional arithmetic operations. This novel kernel adaptive filtering algorithm is applied to recurrent network training, chaotic time-series estimation and cooperative filtering using Gaussian and non-Gaussian noises, and inverse kinematics modeling. Simulation results show FBF outperforms existing Kalman-based algorithms.
\end{abstract}

\end{frontmatter}
\section{Introduction}
The famed Kalman filter \cite{Kalman} produces exact Bayesian inference in a linear dynamical system, when all latent and observed variable distributions are Gaussian. Kalman filters, predictors, and smoothers are enormously successful with a rich array of applications. However, its optimality depends on the linear structure of the underlying system, accurate knowledge of the system parameters, and the exact statistics of the Gaussian noises. In this paper, we address these three shortcomings of the classic Kalman filter with the formulation of a Bayesian recursive state estimator in reproducing kernel Hilbert space (RKHS).

Real-world applications in science and engineering involve nonlinear transformations. The notions of optimality and analytical or general close-form solution become intractable when dealing with such systems \cite{Kushner67SubOptimal}. To overcome this limitation, several suboptimal solutions to the Bayesian filter or nonlinear extensions of the Kalman filter were developed that linearize or perform moment matching to approximate the nonlinear update, such as the extended Kalman filter (EKF) \cite{OptimalFiltering,Haykin01KFNN}, unscented Kalman filter (UKF) \cite{Julier97anew,Wan00theunscented}, and the cubature Kalman filter (CKF) \cite{Arasaratnam09CKF}. The EKF approximates a nonlinear system using first-order linearizion. The UKF uses unscented transform for approximation. The CKF uses a third-degree spherical-radial cubature rule to compute the second-order statistics of a nonlinearly transformed Gaussian random variable \cite{Arasaratnam09CKF}. Recently, several formulations were proposed, using the theory of RKHS, including the kernel Kalman filter (KKF) \cite{Ralaivola03KKF}, the dynamical system model with a conditional embedding (DSMCE) operator \cite{Song09DSMCE}, and the kernel Bayes’ rule (KBR) \cite{Fukumizu13KBR}. The KKF is implemented in a high dimensional subspace obtained by the Kernel principal component analysis (KPCA) algorithm \cite{Scholkopf98KPCA}. The DSMCE and KBR algorithms are both developed based on the embedding of conditional distributions in RKHS.

All of these generative approaches treat the time series or their feature-space mappings as the hidden states and describe the dynamics by the assumed state-space model (SSM) or the given hidden state training data. This brings us to the second major shortcoming of the Kalman filter and its nonlinear extensions, mentioned above. They assume known system dynamics, while most real systems have unknown transformations. Therefore, accurate estimation and prediction cannot be obtained by these algorithms. The kernel Kalman filter based on the conditional embedding operator (KKF-CEO) \cite{Zhu14CEO} was developed to combat this issue. It constructs a state space model in an RKHS using the estimated conditional embedding operator and implements the Kalman filtering in this space. However, similar to the extended KRLS (Ex-KRLS) algorithm \cite{EKRLS}, the KKF-CEO is formulated using a simple additive noise model and does not utilize a full state-space representation. More recently, the kernel adaptive autoregressive-moving-average (ARMA) or KAARMA algorithm \cite{KAARMA} was introduced to bring together the theories of adaptive signal processing and recurrent neural networks (RNNs), extending the current theory of feedforward kernel adaptive filtering to include feedback. It is a true infinite impulse response (IIR) system in the RKHS, formulated with a full SSM. It learns unknown general nonlinear continuous-time state-transition and measurement equations, using only an input sequence and the observed outputs. KAARMA is trained using stochastic gradient-descent and can operate on incomplete or deferred outputs for sequence learning. We have successfully applied KAARMA to model flight dynamics of insects \cite{Li17InsectFSM,Li17Insect}, plant growth patterns \cite{Li17STEM}, and speech using biologically-inspired spike train data \cite{Li2018}.

 In this paper, we derive a Kalman-like filter in the RKHS on the full state-space representation used in KAARMA. This is similar to training RNNs using the extended Kalman filter \cite{Williams92EKFRNN}, except that the network is implemented in a functional space where classical linear methods are used (computed in the input space using the representer theorem and kernel trick), and the universal approximation property of kernel method provides general nonlinear solutions in the input space. To distinguish our work from previous attempts, we name our novel algorithm the Functional Bayesian Filter (FBF). We will show that FBF generalizes the KAARMA algorithm. 

Lastly, the Kalman filter is optimal only for Gaussian random variables (limited to first and second order statistics). This assumption results in degraded performances when data contains outliers or are otherwise non-Gaussian (higher order statistics). The FBF, on the other hand, makes no assumption on the state transition model or the noise profile, by learning directly from measured data. In fact, reproducing kernels are covariance functions explains their early role in inference problems \cite{Aronszajn1950,Parzen1959}. Because of the nonlinearity of the Gaussian kernel, all even moments of the random variable contribute to the estimation of similarity measure \cite{Santamaria2006,LiuCorrentropy2007}. The functional Bayesian filter in the RKHS propagates and updates the full statistics of the measurement distribution, not just the mean and covariance as in its input-space counterpart, which results in enhanced estimation in non-Gaussian noise environments. From an information theoretic learning (ITL) \cite{ITL} perspective, the FBF inherently computes information quantities such as correntropy and information potential to evolve the probability density function (pdf) of the data, using adaptive Parzen estimation. The versatility of the proposed FBF, underthe powerful unifying framework of kernel methods, will be useful for a diverse set of applications in automatic control, machine learning, and signal processing. The major attributes of different well-known Kalman-based Bayesian filters are summarized in Table \ref{tab:Attributes}. 

\begin{table}[ht]\renewcommand{\arraystretch}{1.4}
	\setlength{\tabcolsep}{4pt}
	\scriptsize
	\centering\caption{Comparison of Kalman-based Bayesian filters and FBF.}
	\label{tab:Attributes}
	\begin{tabular}{ |c|c|c|c|}
		\hline		
		\diagbox[width=7.5em]{Filter}{Property} &  \thead{Nonlinear\\Dynamics} &  \thead{Non-Gaussian\\Statistics} &  \thead{Unknown$^*$\\ Dynamics} \\ \hline
		Kalman& \Cross & \Cross & \Cross \\ \hline
		EKF& limited & \Cross & \Cross \\ \hline
		UKF& \Checkmark & \Cross & \Cross \\ \hline
		CKF& \Checkmark & \Cross & \Cross \\ \hline
		FBF& \Checkmark & \Checkmark & \Checkmark \\ \hline
	\end{tabular}
	\normalsize
\end{table}

$^*$Conventional Kalman-based methods require accurate knowledge of the system dynamics. For dual estimation (estimating the state of a dynamic system and the model giving rise to the dynamics), a separate construct such as a neural network can be used as the functional form of the unknown model. This typically requires separate state-space representations for the signal and the weights, i.e., two dynamical systems, one for the evolution of the states, and the other, the evolution of the network parameters. FBF is inherently an ARMA model, using joint estimation (state and model parameters are concatenated within a combined state vector, resulting in a single dynamical model to estimate both quantities simultaneously). There is no \textit{a priori} requirement on system knowledge, as it is trained from scratch using only observations (to take advantage of FBF, the nonlinear SSM has to be expressed in terms of a linear filter in the RKHS, and the simplest conversion is through training). If accurate knowledge of the system is available (i.e., clean states), this will help with convergence during training, but it is not a hard requirement.

The rest of the paper is organized as follows. In Section 2, the Hilbert space SSM is introduced. In Section 3, the proposed FBF algorithm is derived. Then experimental results are presented to compare our novel algorithm with several existing algorithms in Section 4. Finally, Section 5 concludes this paper.

\section{State Space Representation in RKHS}
\begin{figure}[t]
	\centering
	\includegraphics[width=0.48\textwidth]{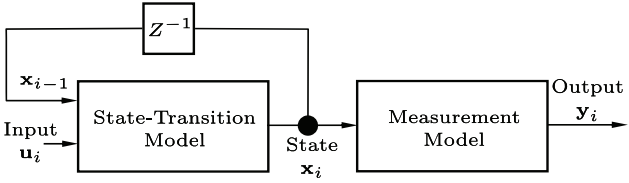}
	\caption{General state-space model for dynamical system.}
	\label{fig:model}
\end{figure}

Let a dynamical system (Fig. \ref{fig:model}) be defined in terms of a general continuous nonlinear state transition and observation functions, $\textbf{f}(\cdot,\cdot)$ and $\textbf{h}(\cdot)$, respectively, 
\begin{align}
\textbf{x}_{i} &= \textbf{f}(\textbf{x}_{i-1},\textbf{u}_{i})\label{ssm1}\\
\textbf{y}_i &= \textbf{h}(\textbf{x}_{i})\label{ssm2}
\end{align}
where 
\begin{align}
\textbf{f}(\textbf{x}_{i-1},\textbf{u}_i) &\stackrel{\Delta}{=} \left[f^{(1)}(\textbf{x}_{i-1},\textbf{u}_i),\cdots,f^{(n_x)}(\textbf{x}_{i-1},\textbf{u}_i)\right]^T\nonumber\\
&=\left[\textbf{x}^{(1)}_i,\cdots,\textbf{x}^{(n_x)}_i\right]^T\\
\textbf{h}(\textbf{x}_i) &\stackrel{\Delta}{=} \left[h^{(1)}(\textbf{x}_i),\cdots,h^{(n_y)}(\textbf{x}_i)\right]^T\nonumber\\
&=\left[\textbf{y}^{(1)}_i,\cdots,\textbf{y}^{(n_y)}_i\right]^T
\end{align}
with input $\textbf{u}_i\in\mathbb{R}^{n_u}$, state $\textbf{x}_i\in\mathbb{R}^{n_x}$, output $\textbf{y}_i\in\mathbb{R}^{n_y}$, and the parenthesized superscript $^{(k)}$ indicating the $k$-th component of a vector or the $k$-th column of a matrix. Note that the input, state, and output vectors have independent degrees of freedom or dimensionality.

For simplicity, we rewrite (\ref{ssm1}-\ref{ssm2}) in terms of a new hidden state vector
\begin{align}
\textbf{s}_{i} &\stackrel{\Delta}{=}
\begin{bmatrix}
\textbf{x}_{i} \\[0.3em]
\textbf{y}_i\\[0.3em]
\end{bmatrix} = \begin{bmatrix}
\textbf{f}(\textbf{x}_{i-1},\textbf{u}_{i})\\[0.3em]
\textbf{h}\circ\textbf{f}(\textbf{x}_{i-1},\textbf{u}_{i})\\[0.3em]
\end{bmatrix} \label{eq:aug_state}\\
\textbf{y}_i &= \textbf{s}_{i}^{(n_s-n_y+1:n_s)} =  \underbrace{\begin{bmatrix}
	\textbf{0} & \textbf{I}_{n_y} \\[0.3em]
	\end{bmatrix}}_{\mathds{I}}\begin{bmatrix}
\textbf{x}_{i} \\[0.3em]
\textbf{y}_i\\[0.3em]
\end{bmatrix}\label{eq:selector}
\end{align}
where $\textbf{I}_{n_y} $ is an $n_y\times n_y$ identity matrix, \textbf{0} is an $n_y\times n_x$ zero matrix, and $\circ$ is the function composition operator. This augmented state vector $\textbf{s}_i\in\mathbb{R}^{n_s}$ is formed by concatenating the output $\textbf{y}_i$ with the original state vector $\textbf{x}_i$. With this rewriting, measurement equation simplifies to a fixed selector matrix $\mathds{I}\stackrel{\Delta}{=} \begin{bmatrix}
\textbf{0} & \textbf{I}_{n_y} \\[0.3em]
\end{bmatrix}$. Note, despite the parsimonious structure of (\ref{eq:selector}), there is no restriction on the measurement equation, as $\textbf{h}\circ\textbf{f}$ in (\ref{eq:aug_state}) is its own set of general nonlinear equations, i.e., this is equivalent to the generative model shown in Fig. \ref{fig:model}. 

Next, we define an equivalent transition function $\textbf{g}(\textbf{s}_{i-1},\textbf{u}_{i})=\textbf{f}(\textbf{x}_{i-1},\textbf{u}_{i})$ taking as argument the new state variable $\textbf{s}$. Using this notation, (\ref{ssm1}-\ref{ssm2}) becomes
\begin{align}
\textbf{x}_{i} &= \textbf{g}(\textbf{s}_{i-1},\textbf{u}_{i})\label{nssm1}\\
\textbf{y}_i &= \textbf{h}(\textbf{x}_{i})\label{nssm2} = \textbf{h}\circ\textbf{g}(\textbf{s}_{i-1},\textbf{u}_{i}).
\end{align}

To learn the general continuous nonlinear transition and observation functions, $\textbf{g}(\cdot,\cdot)$ and $\textbf{h}\circ \textbf{g}(\cdot,\cdot)$, respectively,  we apply the theory of RKHS. First, we map the augmented state vector $\textbf{s}_{i}$ and the input vector $\textbf{u}_{i}$ into two separate RKHSs as $\varphi(\textbf{s}_{i})\in\mathcal{H}_s$ and $\phi(\textbf{u}_{i})\in\mathcal{H}_u$, respectively.  By the representer theorem, the state-space model defined by (\ref{nssm1}-\ref{nssm2}) can be expressed as the following set of weights (functions in the input space) in the joint RKHS  $\mathcal{H}_{su}\stackrel{\Delta}{=}\mathcal{H}_{s}\otimes\mathcal{H}_{u}$ 
\begin{align}
{\bm\Omega}\stackrel{\Delta}{=}{\bm\Omega}_{\mathcal{H}_{su}}\stackrel{\Delta}{=} \begin{bmatrix}
\textbf{g}(\cdot,\cdot)\\[0.3em]
\textbf{h}\circ \textbf{g}(\cdot,\cdot) \\[0.3em]\end{bmatrix}
\end{align}
where $\otimes$ is the tensor-product operator. We define the new features in the tensor-product RKHS as
\begin{align} \psi(\textbf{s}_{i-1},\textbf{u}_{i})\stackrel{\Delta}{=}\varphi(\textbf{s}_{i-1})\otimes\phi(\textbf{u}_i)\in \mathcal{H}_{su}.
\end{align} 
It follows that the tensor-product kernel is defined by
\begin{align}
\langle\psi(\textbf{s},\textbf{u}),\psi(\textbf{s}',\textbf{u}')\rangle_{\mathcal{H}_{su}}&\stackrel{\Delta}{=}\mathcal{K}_{su}(\textbf{s},\textbf{u},\textbf{s}',\textbf{u}')\nonumber\\
&=(\mathcal{K}_{s}\otimes\mathcal{K}_{u})(\textbf{s},\textbf{u},\textbf{s}',\textbf{u}')\nonumber\\
&=\mathcal{K}_{s}(\textbf{s},\textbf{s}')\cdot\mathcal{K}_{u}(\textbf{u},\textbf{u}').\label{TPkernel}
\end{align}
This construction has several advantages over the simple concatenation of the input $\textbf{u}$ and the state $\textbf{s}$. First, the tensor product kernel of two positive definite kernels is also a positive definite kernel \cite{Scholkopf01kernel}. Second, since the adaptive filtering is performed in an RKHS using features, there is no constraint on the original input signals or the number of signals, as long as we use the appropriate reproducing kernel for each signal. Last but not least, this formulation imposes no restriction on the relationship between the signals in the original input space. This is important for input signals having different representations and spatio-temporal scales. 

\begin{figure}[h]
	\centering
	\includegraphics[width=0.48\textwidth]{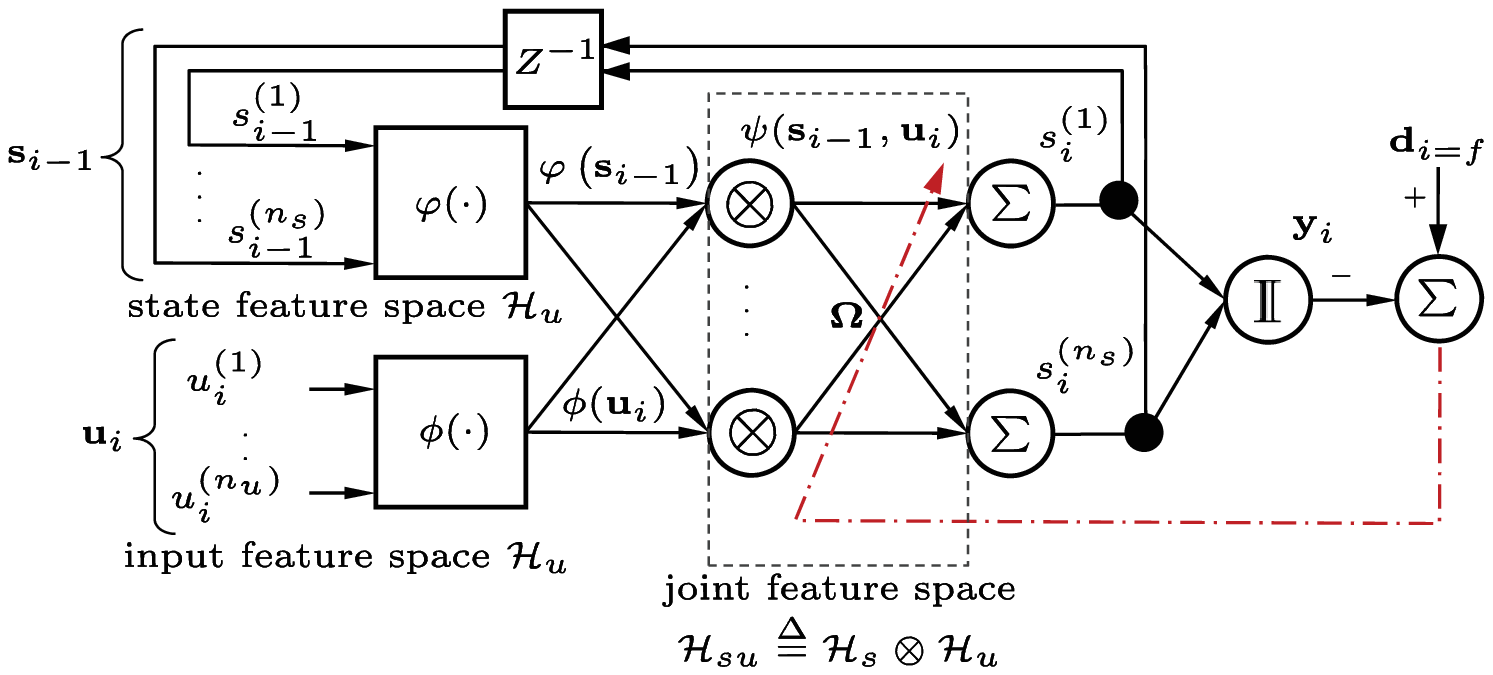}
	\caption{ARMA model in the RKHS.}
	\label{fig:model4}
\end{figure}

Finally, the kernel state-space model becomes
\begin{align}
\textbf{s}_{i}&={\bm\Omega}^T\psi(\textbf{s}_{i-1},\textbf{u}_{i})\label{algGenerateState}\\
\textbf{y}_i&=\mathds{I}\textbf{s}_i.\label{algOutput}
\end{align}
Fig. \ref{fig:model4} shows a simple kernel ARMA model corresponding to the SSM in (\ref{algGenerateState} and \ref{algOutput}). In general, the states $\textbf{s}_i$ are assumed hidden, and the desired does not need to be available at every time step, e.g., a deferred desired output value for $\textbf{y}_i$ may only be observed at the final indexed step $i=f$, i.e., $\textbf{d}_f$.

\section{Functional Bayesian Filter}
Given the history or sequence of input-observations $\textbf{Y}_{i-1} = \{\textbf{y}_j\}^{(i-1)}_{j=1}$ up to time index $(i-1)$, the objective is to estimate the state $\textbf{x}_i$. Let $\hat{\textbf{x}}_{i|i-1}$ be the \textit{a priori} state estimate at step $i$, given knowledge of the process prior to step $i$, and $\hat{\textbf{x}}_{i|i}$ be the \textit{a posteriori} state estimate at step $i$, given new measurement $\textbf{y}_i$. We can define \textit{a priori} and \textit{a posteriori} estimate errors as
\begin{align}
\textbf{e}_{i|i-1} &= \textbf{x}_i-\hat{\textbf{x}}_{i|i-1}\\
\textbf{e}_{i|i} &= \textbf{x}_i-\hat{\textbf{x}}_{i|i}.
\end{align}
The goal is to minimize the a posteriori error covariance 
\begin{align}
\textbf{P}_{i|i} = E[\textbf{e}_{i|i}\textbf{e}^\intercal_{i|i}].
\end{align}
This process consists of two distinct phases. The \textit{time update} computes the predictive density
\begin{align}
p(\textbf{x}_i|\textbf{Y}_{i-1})=\int_{\mathbb{R}^{n_x}}{p(\textbf{x}_i|\textbf{x}_{i-1})p(\textbf{x}_{i-1}|\textbf{Y}_{i-1})d_{\textbf{x}_{i-1}}}.
\end{align}
The \textit{measurement update} involves computing the posterior density of the current state which is proportional to the product of the measurement likelihood and the predicted state (Bayes' rule) as
\begin{align}
p(\textbf{x}_i|\textbf{Y}_{i})=\frac{p(\textbf{y}_i|\textbf{x}_{i})p(\textbf{x}_i|\textbf{Y}_{i-1})}{p(\textbf{y}_i|\textbf{Y}_{i-1})}
\end{align}
where the denominator or normalizing constant is
\begin{align}
p(\textbf{y}_i|\textbf{Y}_{i-1}) = \int_{\mathbb{R}^{n_x}}{p(\textbf{y}_i|\textbf{x}_{i})p(\textbf{x}_i|\textbf{Y}_{i-1})d_{\textbf{x}_{i}}}.
\end{align}
The key to Bayesian filtering is the Gaussian assumption: both the predictive density $p(\textbf{x}_i|\textbf{Y}_{i-1})$ and the filter likelihood density $p(\textbf{y}_i|\textbf{x}_i)$ are Gaussian, which results in a Gaussian posterior density $p(\textbf{x}_i|\textbf{Y}_i)$. The time update computes the conditional mean $\hat{\textbf{x}}_{i|i-1}$ and the associated covariance $\textbf{P}_{i|i-1}$ of the Gaussian predictive density as 
\begin{align}
\hat{\textbf{x}}_{i|i-1}&=E[\textbf{x}_i|\textbf{Y}_{i-1}]
\end{align}
and
\begin{align}
\textbf{P}_{i|i-1}= E[(\textbf{x}_i-\hat{\textbf{x}}_{i|i-1})(\textbf{x}_i-\hat{\textbf{x}}_{i|i-1})^\intercal|\textbf{y}_{1:i-1}].
\end{align}
The errors in predicted measurements are zero-mean white sequences with filter likelihood density approximated by the Gaussian as
\begin{align}
p(\textbf{y}_i|\textbf{Y}_{i-1})=\mathcal{N}(\textbf{y}_i;\hat{\textbf{y}}_{i|i-1},\textbf{P}_{yy,i|i-1})
\end{align}
where $\hat{\textbf{y}}_{i|i-1}$ is the predicted measurement with the associated covariance $\textbf{P}_{yy,i|i-1}$. The conditional Gaussian density of the joint state and measurement is
\begin{align}
\scalebox{.9}{$p\left( \begin{bmatrix}
	\textbf{x}_i \\[0.3em]
	\textbf{y}_i\\[0.3em]
	\end{bmatrix}|\textbf{Y}_{i-1}\right)=\mathcal{N}\left(\begin{bmatrix}
	\hat{\textbf{x}}_{i|i-1} \\[0.3em]
	\hat{\textbf{y}}_{i|i-1}\\[0.3em]
	\end{bmatrix},\begin{bmatrix}
	\textbf{P}_{i|i-1} & \textbf{P}_{xy,i|i-1}\\[0.3em]
	\textbf{P}^\intercal_{xy,i|i-1} & \textbf{P}_{yy,i|i-1}\\[0.3em]
	\end{bmatrix}\right)$}
\end{align}
where $\textbf{P}_{xy,i|i-1}$ is the cross-covariance.

On the receipt of a new measurement $\textbf{y}_k$, the Bayesian filter copmutes the posterior density $p(\textbf{x}_k|\textbf{Y}_k)$ yielding
\begin{align}
p(\textbf{x}_i|\textbf{Y}_i) = \mathcal{N}(\textbf{x}_i;\hat{\textbf{x}}_{i|i},\textbf{P}_{i|i})
\end{align}
where
\begin{align}
\hat{\textbf{x}}_{i|i} &= \hat{\textbf{x}}_{i|i-1} + \textbf{G}_i(\textbf{y}_i-\hat{\textbf{y}}_{i|i-1})\\
\textbf{P}_{i|i} &= \textbf{P}_{i|i-1} - \textbf{G}_i\textbf{P}_{yy,i|i-1}\textbf{G}_i^\intercal\\
\textbf{G}_i &= \textbf{P}_{xy,i|i-1}\textbf{P}^{-1}_{yy,i|i-1}
\end{align}
with $\textbf{G}_i$ being the gain or fusion factor that minimizes the \textit{a posteriori} error covariance $\textbf{P}_{i|i}$. While the recursive estimation is linear, no assumption has been made on the linearity of the model. If $f(\cdot)$ and $h(\cdot)$ are linear functions of the state, the Bayesian filter under the Gaussian assumption reduces to the Kalman filter. In such case the mean squared error (MSE) provides the value of $\hat{\textbf{x}_i}$ which maximizes the likelihood of the signal $\textbf{y}_i$. The Kalman filter is optimal in the sense that it minimizes the estimated error covariance. The Gaussian pdf is widely used due to its convenient mathematical properties: closed under linear transformation and conditioning, and uncorrelated jointly Gaussian random variables are independent. Although rarely do the conditions necessary for optimality actually exist, nonetheless the Kalman filter performs well for many applications, due to its relative simplicity and robustness, and the Gaussian pdf approximates physical random phenomena by virtue of the central limit theorem \cite{Arasaratnam09CKF}. 

\subsection{Information theoretic learning (ITL)}
The aesthetics of Kalman filtering lies in its parsimonious form, by considering only the uncertainty represented in the covariance. It is the optimal MSE filter. In developing the functional Bayesian filter, we will see that higher order moments are automatically preserved that gives not only the minimum mean-square error estimate of the state, but also an information theoretic criterion for non-Gaussian uncertainties. Here, we provide a brief background.

ITL is a framework to adapt nonparametric systems using information quantities such as entropy and divergence \cite{ITL}. ITL criteria is still directly estimated from data via Parzen kernel estimator, but it extracts more information from the data for adaptation, and yields, therefore, solutions that are more accurate than MSE in non-Gaussian and nonlinear signal processing. Reproducing kernels are covariance functions explains their early role in inference problems \cite{Aronszajn1950,Parzen1959}. Renyi's quadratic entropy of a random variable $X$ with pdf $f_X(x)$ is defined as
\begin{align}
H_2(X) \stackrel{\Delta}{=} -\log\int f^2_X(x)dx.
\end{align}
The Parzen estimate of the pdf, given a set of independent and identically distributed (i.i.d.) data ${x_i}^N_{i=1}$ drawn from the distribution is 
\begin{align}
\hat{f}_{X;\sigma}(x)=\frac{1}{N}\sum^{N}_{i=1}\mathcal{K}_\sigma(x-x_i)
\end{align}
where $N$ is the number of data points and $\mathcal{K}_\sigma$ is the Gaussian kernel with kernel size $\sigma$
\begin{align}
\mathcal{K}_\sigma(x-x_i)=\frac{1}{\sqrt{2\pi}\sigma}\exp\left(-\frac{(x-x_i)^2}{2\sigma^2}\right).
\end{align}
Without loss of generality, we will only consider the Gaussian kernel in this paper.

A nonparametric estimate of Renyi's quadratic entropy directly from samples is
\begin{align}
\hat{H}_2(X)=-\log {\rm IP}(X)
\end{align}
where the information potential (IP) is defined as
\begin{align}
{\rm IP}(X)\stackrel{\Delta}{=}\frac{1}{N^2}\sum^{N}_{i=1}\sum^{N}_{j=1}\mathcal{K}_{\sqrt{2}\sigma}(x_i-x_j).
\end{align}

Let ${X_t,t\in T}$ be a stochastic process with $T$ being an index set. The nonlinear mapping $\Phi$ induced by the Gaussian kernel maps the data into the feature space $\mathbb{F}$, where the auto-correntropy function $V_X(t,t+\tau)$ is defined from $T\times T$ into $\mathbb{R}^+$ given by
\begin{align}
V_X(t,t+\tau)&\stackrel{\Delta}{=}\mathrm{E}[\langle\Phi(X_t),\Phi(X_{t+\tau})\rangle_\mathbb{F}]\\
&=\mathrm{E}[\mathcal{K}_{\sigma}(X_t-X_{t+\tau})].
\end{align}
where $\mathrm{E}[\cdot]$ denotes the expectation. A sufficient condition for $V(t,t-\tau) = V(\tau)$ is that the stochastic process must be strictly stationary on all the even moments, a stronger condition than wide sense stationarity (limited to \nth{2} order moments). The IP is the mean squared projected data $\langle \frac{1}{N}\sum^N_{i=1}\Phi(x_i),\frac{1}{N}\sum^N_{j                                                                                                                                                                                   =1}\Phi(x_j) \rangle$or the expected value of correntropy over lags $\tau$. A more general form of correntropy (cross-correntropy) between two random variables is defined as
\begin{align}
V_\sigma(X,Y) \stackrel{\Delta}{=} \mathrm{E}[\mathcal{K}_{\sigma}(X-Y)].
\end{align}
The sample estimate of correntropy for a finite number of data ${(x_i,y_i)}^N_{i=1}$ is
\begin{align}
\hat{V}_{N,\sigma}(X,Y)=\frac{1}{N}\sum^N_{i=1}\mathcal{K}_{\sigma}(x_i-y_i).
\end{align}
Using Taylor series expansion for the Gaussian kernel, correntropy can be expressed as
\begin{align}
V_{\sigma}(X,Y)=\frac{1}{\sqrt{2\pi}\sigma}\sum^\infty_{i=0}\frac{(-1)^n}{2^n\sigma^{2n}n!}\mathrm{E}[(X-Y)^{2n}]
\end{align}
which involves all the even-order moments of the random variable $X-Y$ (where kernel choice dictates the expansion, e.g., the sigmoidal kernel contains all the odd moments) \cite{Santamaria2006}. 

In fact, all learning algorithms that use nonparametric pdf estimates in the input space admit an alternative formulation as kernel methods expressed in terms of inner products. As shown above, the kernel techniques are able to extract higher order statistics of the data that should lead to performance improvements for non-Gaussian environments. A major limitation of conventional statistical measures is the i.i.d. assumption. Most practical problems, however, involve some correlation or temporal structure. Therefore, most are not using all the available information in the case of temporally correlated (non-white) input signals. Unlike conventional measures, the generalized correlation function effectively exploits both the statistical and the temporal information about the input signal. We will see in the following section that this feature is intrinsic in the functional Bayesian filtering, even without explicating defining ITL cost functions. 
\subsection{Bayesian filtering in the RKHS}
Let the discrete-time dynamical system be described by the following state transition equation and measurement equation:
\begin{align}
{\textbf{x}}_i &= \textbf{F}_{i-1}{\textbf{x}}_{i-1} + \textbf{w}_{i-1} \label{eq:StateTrans}\\
{\textbf{y}}_i &= \textbf{H}_i{\textbf{x}}_i + \textbf{v}_i.\label{eq:Meas}
\end{align}

For the ARMA model in the RKHS shown in Fig. \ref{fig:model4}, we define the state as the following super augmented vector
\begin{align}
\textbf{x}'_i \stackrel{\Delta}{=} \begin{bmatrix}
\textbf{s}_i \\[0.3em]
\bm{\Omega}_i\\[0.3em]
\end{bmatrix}\label{Eq:AugState}
\end{align}
where $\textbf{s}_{i} =[\textbf{x}_{i},  \textbf{y}_i]^\intercal$, same as in (\ref{eq:aug_state}), and we treat the weight matrix $\bm{\Omega}_i$ in the RKHS at time $i$ as an $n_{\bm{\Omega}}$-dimensional vector rather than a matrix, by grouping the weight parameters in an orderly fashion. The state transition matrix can be expressed in block form as
\begin{align}
\textbf{F} = \begin{bmatrix}
\textbf{F}_1 & \textbf{F}_2 \\[0.3em]
\textbf{0} & \textbf{I}_{n_{\bm{\Omega}}} \\[0.3em]
\end{bmatrix} \label{eq:F}
\end{align}
where $\textbf{F}_1$ is an $n_s\times n_s$ matrix, $\textbf{F}_2$ is an $n_s\times n_{\bm{\Omega}}$ matrix, and $\textbf{I}_{n_{\bm{\Omega}}}$ is an $n_{\bm{\Omega}}\times n_{\bm{\Omega}}$ identity matrix. 
State transition equation (\ref{eq:StateTrans}) becomes
\begin{align}
\begin{bmatrix}
\textbf{s}_i \\[0.3em]
\bm{\Omega}_i\\[0.3em]
\end{bmatrix} = \begin{bmatrix}
\textbf{F}_1 & \textbf{F}_2 \\[0.3em]
\textbf{0} & \textbf{I}_{n_{\bm{\Omega}}} \\[0.3em]
\end{bmatrix}\begin{bmatrix}
\textbf{s}_{i-1} \\[0.3em]
\bm{\Omega}_i\\[0.3em]
\end{bmatrix} + \textbf{w}_{i-1} \label{Eq:SSM}
\end{align}
where we assume that the weights in the RKHS observe the trivial dynamics $\bm{\Omega}_i =  \bm{\Omega}_{i-1}$. Since the network output $\textbf{y}_i$ is a subvector of the hidden state $\textbf{s}_{i}$, the measurement function $\textbf{H}_i$ in (\ref{eq:Meas}) becomes a simple projection onto the last $n_y$ components of $\textbf{s}_i$
\begin{align}
\textbf{y}_i &= \textbf{H} \begin{bmatrix}
\textbf{s}_i \\[0.3em]
\bm{\Omega}_i\\[0.3em]
\end{bmatrix} + \textbf{v}_i
\end{align}
where
\begin{align}
\textbf{H} \stackrel{\Delta}{=} \begin{bmatrix} \mathds{I} & \textbf{0} \end{bmatrix}\label{eq:H}
\end{align}
with $\mathds{I}\stackrel{\Delta}{=} \begin{bmatrix}
\textbf{0} & \textbf{I}_{n_y} \\[0.3em]
\end{bmatrix}\in \mathbb{R}^{n_y\times n_s}$ being a fixed selector matrix.

From the state-space model in (\ref{Eq:SSM}), the state transition matrix blocks in (\ref{eq:F}) are 
\begin{align}
\textbf{F}_1(i) = \frac{\partial \textbf{s}_i}{\partial \textbf{s}_{i-1}}
\end{align}
and
\begin{align}
\textbf{F}_2(i) = \frac{\partial \textbf{s}_i}{\partial \bm{\Omega}_i}.
\end{align}
Using the representer theorem, RKHS weights ${\bm\Omega}_i$ at time $i$ (where $i\geq 1$) can be written as a linear combination of prior features
\begin{align}
{\bm\Omega}_i = {\bm\Psi}_i \textbf{A}_i \label{representation}
\end{align}
where ${\bm\Psi}_i \stackrel{\Delta}{=} \left[\psi(\textbf{s}_{-1},\textbf{u}_0),\cdots,\psi(\textbf{s}_{N-2},\textbf{u}_{N-1})\right]\in\mathbb{R}^{n_{\bm\Omega}\times N}$ is a collection of the $N$ past tensor-product features, and $\textbf{A}_i \stackrel{\Delta}{=} \left[\bm{\alpha}_{i,1},\cdots,\bm{\alpha}_{i,n_s} \right]\in\mathbb{R}^{N\times n_s}$ is the set of coefficients with column vector $\bm{\alpha}_{i,k}\in\mathbb{R}^{N}$ corresponding to the $k$-th state dimension $(1\leq k \leq n_s)$. For conventional kernel adaptive filters, the number of basis functions grows linearly with each new sample, i.e., $N = i$. Here, we use $N\geq 1$ to denote a dictionary ${\bm\Psi}_i$ of arbitrary size, with $\psi(\textbf{s}_{-1},\textbf{u}_0)$ initialization. Thus, each of the $k$-th state component of the filter weights at time $i$ becomes
\begin{align}
{\bm\Omega}^{(k)}_i = {\bm\Psi}_i \textbf{A}^{(k)}_i = {\bm\Psi}_i \bm{\alpha}_{i,k}
\end{align}
which corresponds to a general nonlinear function in the input space.

Since the hidden states are propagated using linear operator in the RKHS, i.e., $\textbf{s}_i = {\bm\Omega}^\intercal_i \psi(\textbf{s}_{i-1},\textbf{u}_{i})$, using the representer theorem (\ref{representation}), we can compute $\textbf{F}_1(i)$ in the input space as
\begin{align}
\frac{\partial \textbf{s}_i}{\partial \textbf{s}_{i-1}} &= \frac{\partial{\bm\Omega}^\intercal_i \psi(\textbf{s}_{i-1},\textbf{u}_{i})}{\partial \textbf{s}_{i-1}}\nonumber\\
&= \textbf{A}^T_i\frac{ \partial{\bm\Psi}^\intercal_i\psi(\textbf{s}_{i-1},\textbf{u}_{i})}{\partial \textbf{s}_{i-1}}\nonumber\\
&= \underbrace{2a_s \textbf{A}^\intercal_i \textbf{K}_i \textbf{D}^\intercal_i}_{{\bm\varLambda}_i} \label{Eq:F1}
\end{align}
where the partial derivative is evaluated using Gaussian tensor-product kernel, and
\begin{align}
\textbf{K}_i \stackrel{\Delta}{=} {\rm diag}({\bm\Psi}_i^\intercal\psi(\textbf{s}_{i-1},\textbf{u}_{i}))
\end{align}
is an $N\times N$ diagonal matrix with eigenvalues $\textbf{K}_i^{(j,j)}=\mathcal{K}_{a_s}(\textbf{s}_j,\textbf{s}_{i-1})\cdot\mathcal{K}_{a_u}(\textbf{u}_j,\textbf{u}_i)$, where $1\leq j \leq N$, and $\textbf{D}_i\stackrel{\Delta}{=}\left[(\textbf{s}_{-1}-\textbf{s}_{i-1}),\cdots,(\textbf{s}_{N-2}-\textbf{s}_{i-1})\right]\in\mathbb{R}^{n_s\times N}$ is the difference matrix between state centers of the filter and the current input state variable $\textbf{s}_{i-1}$. We collect the terms into an $n_s\times n_s$ matrix 
\begin{align}
{\bm\varLambda}_i \stackrel{\Delta}{=} \frac{\partial \textbf{s}_{i}}{\partial \textbf{s}_{i-1}}=2a_s \textbf{A}^\intercal_i \textbf{K}_i \textbf{D}^\intercal_i
\end{align} 
which we call the state-transition gradient, where each entry $(1\leq l, m \leq n_s)$ can be expressed as
\begin{align}
{\bm\varLambda}^{(l,m)}_i &=2a_s\sum^N_{k=1}\underbrace{\bm{\alpha}^{(k)}_{i,l}\left(\textbf{s}^{(m)}_{k-2}-\textbf{s}^{(m)}_{i-1}\right)}_{\mathbbm{a}^{k,m}_{i,l}}\textbf{K}_i^{(k,k)}\nonumber\\
&=2a_s\sum^N_{k=1}\mathbbm{a}^{k,m}_{i,l}\mathcal{K}_{a_s}(\textbf{s}_k,\textbf{s}_{i-1})\cdot\mathcal{K}_{a_u}(\textbf{u}_k,\textbf{u}_i)
\end{align}
which has an information theoretic interpretation as a weighted Parzen estimate of the joint input data  $(\textbf{s}_i,\textbf{u}_i)$ pdf.

The second block is computed as
\begin{align}
\textbf{F}_2(i) =  \textbf{1}_{n_s}\psi(\textbf{s}_{i-1},\textbf{u}_{i})^\intercal \label{Eq:F2}
\end{align}
where $\textbf{1}_{n_s}\in\mathbf{R}^{n_s\times 1}$ is a vector of all ones.

The estimated state covariance matrix is given by
\begin{align}
\textbf{P}_{i|i-1} = \textbf{F}_i\textbf{P}_{i-1|i-1}\textbf{F}_i^\intercal+\textbf{Q}_{i-1}\label{eq:CovEst}
\end{align}
where 
\begin{align}
\textbf{Q}_i =\mathbf{E}\left[\textbf{w}_i \textbf{w}^\intercal_i\right] 
\end{align}
is the process noise covariance matrix. The block structure of the state transition matrix $\textbf{F}$ leads to the following decomposition in the estimated state covariance matrix
\begin{align}
\textbf{P} = \begin{bmatrix}
\textbf{P}_1 & \textbf{P}_2 \\[0.3em]
\textbf{P}_3 & \textbf{P}_4 \\[0.3em]
\end{bmatrix}\label{eq:P}
\end{align}
where $\textbf{P} \in \mathbb{R}^{(n_s+n_{\bm{\Omega}})\times (n_s+n_{\bm{\Omega}})}$, $\textbf{P}_1 \in \mathbb{R}^{n_s\times n_s}$, and $\textbf{P}_4\in \mathbb{R}^{n_{\bm{\Omega}}\times n_{\bm{\Omega}}}$ are symmetric, with $\textbf{P}_2 \in \mathbb{R}^{n_s\times n_{\bm{\Omega}}} = \textbf{P}_3^\intercal$. Substituting \ref{eq:P} and \ref{eq:F} into (\ref{eq:CovEst}) yields
\begin{eqnarray}
\lefteqn{\begin{bmatrix}
\textbf{P}_1(i|i-1) & \textbf{P}_2(i|i-1) \\[0.3em]
\textbf{P}_3(i|i-1) & \textbf{P}_4(i|i-1) \\[0.3em]
\end{bmatrix} =} \nonumber\\
& & \scalebox{.9}{$\begin{bmatrix}
\textbf{F}_1(i) & \textbf{F}_2(i) \\[0.3em]
\textbf{0} & \textbf{I}_{n_\Omega} \\[0.3em]
\end{bmatrix}	\begin{bmatrix}
\textbf{P}_1(i-1|i-1) & \textbf{P}_2(i-1|i-1) \\[0.3em]
\textbf{P}_3(i-1|i-1) & \textbf{P}_4(i-1|i-1) \\[0.3em]
\end{bmatrix}\begin{bmatrix}
\textbf{F}^\intercal_1(i) & \textbf{0} \\[0.3em]
\textbf{F}^\intercal_2(i) & \textbf{I}_{n_\Omega} \\[0.3em]
\end{bmatrix}$} \nonumber\\
& & \scalebox{.9}{$+ \begin{bmatrix}
\sigma^2_{s}\textbf{I}_{n_s} & \textbf{0} \\[0.3em]
\textbf{0} & \sigma^2_{\Omega}\textbf{I}_{n_\Omega} \\[0.3em]
\end{bmatrix}$}.
\end{eqnarray}
Using the superscripts $^-$ and $^+$ as shorthands for the \textit{a priori} estimate $(i|i-1)$ and \textit{a posteriori} estimates $(i-1|i-i)$ or $(i|i)$, where appropriate, we obtain the following update rules:
\begin{align}
\textbf{P}^{-}_1 &= \scalebox{.85}{$\left[\textbf{F}_1\textbf{P}^+_1 + \textbf{F}_2(\textbf{P}^+_2)^\intercal\right]\textbf{F}^\intercal_1 + \underbrace{\left[\textbf{F}_1\textbf{P}^{+}_2+\textbf{F}_2\textbf{P}^{+}_4\right]}_{\textbf{P}_2^-} \textbf{F}_2^\intercal+\sigma^2_U \textbf{I}_{n_s}\label{Eq:P1Min}$}\\
\textbf{P}^{-}_2 &= \textbf{F}_1\textbf{P}^{+}_2+\textbf{F}_2\textbf{P}^{+}_4\label{eq:P2}\\
\textbf{P}^{-}_3 &= \textbf{P}^{-\intercal}_2 \\
\textbf{P}^{-}_4 &= \textbf{P}^{+}_4+\sigma^2_\Omega \textbf{I}_{n_\psi}\label{Eq:P4Min}.
\end{align}

The Kalman gain $\mathds{K}$ is computed using the innovation covariance
\begin{align}
\mathds{S}_i &= \textbf{H} \textbf{P}^-\textbf{H}^\intercal + \textbf{R}\nonumber\\
&= \textbf{H} \textbf{P}^-\textbf{H}^\intercal + \sigma_y^2\textbf{I}_{n_y}
\end{align}
where $\textbf{R}$ and $\sigma_y$ are the measurement covariance matrix and output standard deviation, respectively, yielding
\begin{align}
\mathds{K}_i &= \textbf{P}^-\textbf{H}_i^\intercal \mathds{S}_i^{-1}\nonumber\\
&=	\scalebox{.9}{$\begin{bmatrix}
\textbf{P}_1^- & \textbf{P}_2^- \\[0.3em]
\textbf{P}_3^- & \textbf{P}_4^- \\[0.3em]
\end{bmatrix} \begin{bmatrix}
\mathds{I}^\intercal \\[0.3em]
\textbf{0} \\[0.3em]
\end{bmatrix}\left(\begin{bmatrix}
\mathds{I} & \textbf{0}\\[0.3em]
\end{bmatrix}\begin{bmatrix}
\textbf{P}_1^- & \textbf{P}_2^- \\[0.3em]
\textbf{P}_3^- & \textbf{P}_4^- \\[0.3em]
\end{bmatrix} \begin{bmatrix}
\mathds{I}^\intercal \\[0.3em]
\textbf{0} \\[0.3em]
\end{bmatrix} + \sigma_y^2\textbf{I}_{n_y} \right)^{-1}$}\nonumber\\
&= \begin{bmatrix}
\textbf{P}_1^-\mathds{I}^\intercal \\[0.3em]
\textbf{P}_3^-\mathds{I}^\intercal \\[0.3em]
\end{bmatrix}\left(\mathds{I}\textbf{P}_1^-
\mathds{I}^\intercal + \sigma_y^2\textbf{I}_{n_y} \right)^{-1}\nonumber\\
&= \begin{bmatrix}
\textbf{P}_1^-\mathds{I}^\intercal \\[0.3em]
\left(\textbf{P}_2^-\right)^\intercal\mathds{I}^\intercal \\[0.3em]
\end{bmatrix}\left(\mathds{I}\textbf{P}_1^-
\mathds{I}^\intercal + \sigma_y^2\textbf{I}_{n_y} \right)^{-1}\nonumber\\
&\stackrel{\Delta}{=} \begin{bmatrix}
\textbf{L}_1 \\[0.3em]
\textbf{L}_2 \\[0.3em]
\end{bmatrix}\left(\textbf{M}_i + \sigma_y^2\textbf{I}_{n_y} \right)^{-1}\nonumber\\
&\stackrel{\Delta}{=} \begin{bmatrix}
\textbf{L}_1 \\[0.3em]
\textbf{L}_2 \\[0.3em]
\end{bmatrix}\textbf{N}_i
\end{align}
where in order to clean up the notation, we defined the following intermediate matrices
\begin{align}
\textbf{L}_1 &\stackrel{\Delta}{=} \textbf{P}^-_1\mathds{I}^\intercal\label{Eq:L1}\\
\textbf{L}_2 &\stackrel{\Delta}{=} (\textbf{P}^-_2)^\intercal\mathds{I}^\intercal\label{Eq:L2}\\
\textbf{M} &\stackrel{\Delta}{=} \mathds{I}\textbf{L}_1 \label{Eq:M}\\
\textbf{N} &\stackrel{\Delta}{=} (\textbf{M}+\sigma^2_y \textbf{I}_{n_y})^{-1}\label{Eq:N}
\end{align}
with $\textbf{L}_1\in\mathbb{R}^{n_s\times n_y}$ (last $n_y$ columns of $\textbf{P}_1^-$), $\textbf{L}_2\in\mathbb{R}^{n_{\bm{\Omega}}\times n_y}$ (last $n_y$ rows of $\textbf{P}_2^-$, transposed), and $\textbf{M}_i\in\mathbb{R}^{n_y\times n_y}$ (the $n_y\times n_y$ lower-right corner of $\textbf{P}_1^-$) being submatrices of the decomposed state-covariance matrix $\textbf{P}$, since the linear mappings $\textbf{H}$ and $\mathds{I}$ are defined in (\ref{eq:H}) as simple projections onto the last $n_y$ coordinates of state $\textbf{s}_i$, and $\textbf{N}$ is the inverse of the innovation covariance matrix.

Clearly, Kalman gain matrix is decomposed into two components
\begin{align}
\mathds{K}_i &= \begin{bmatrix}
\mathds{K}_1 \\[0.3em]
\mathds{K}_2 \\[0.3em]
\end{bmatrix}=\begin{bmatrix}
\textbf{L}_1\textbf{N}_i \\[0.3em]
\textbf{L}_2\textbf{N}_i \\[0.3em]
\end{bmatrix}\label{Eq:K}
\end{align}
where $\mathds{K}_1\in\mathbb{R}^{n_s\times n_y}$ describes the changes in network activity (states $s_i$), and $\mathds{K}_2\in\mathbb{R}^{n_{\bm{\Omega}}\times n_y}$ corresponds to weight $\bm{\Omega}_i$ changes in response to errors. 

Updating the \textit{a posteriori} state estimate gives
\begin{align}
\textbf{x}^+ &= \textbf{x}^- + \mathds{K}_i\textbf{e}_i\nonumber\\
\begin{bmatrix}
\textbf{s} \\[0.3em]
\bm{\Omega}\\[0.3em]
\end{bmatrix}^+ &=\begin{bmatrix}
\textbf{s} \\[0.3em]
\bm{\Omega}\\[0.3em]
\end{bmatrix}^- +
\begin{bmatrix}
\mathds{K}_1 \\[0.3em]
\mathds{K}_2 \\[0.3em]
\end{bmatrix}\textbf{e}_i.
\end{align}

Updating the \textit{a posteriori} covariance estimate gives
\begin{align}
\textbf{P}^+ &= (\textbf{I} - \mathds{K}\textbf{H})\textbf{P}^-\nonumber\\
\begin{bmatrix}
\textbf{P}_1^+ & \textbf{P}_2^+ \\[0.3em]
\textbf{P}_3^+ & \textbf{P}_4^+ \\[0.3em]
\end{bmatrix} &= \left(\textbf{I} - \begin{bmatrix}
\mathds{K}_1 \\[0.3em]
\mathds{K}_2 \\[0.3em]
\end{bmatrix}\begin{bmatrix}
\mathds{I} & \textbf{0}\\[0.3em]
\end{bmatrix}\right)\begin{bmatrix}
\textbf{P}_1^- & \textbf{P}_2^- \\[0.3em]
\textbf{P}_3^- & \textbf{P}_4^- \\[0.3em]
\end{bmatrix} \nonumber\nonumber\\
&=\begin{bmatrix}
\textbf{P}_1^- & \textbf{P}_2^- \\[0.3em]
\textbf{P}_3^- & \textbf{P}_4^- \\[0.3em]
\end{bmatrix} - \begin{bmatrix}
\mathds{K}_1 \\[0.3em]
\mathds{K}_2 \\[0.3em]
\end{bmatrix}\begin{bmatrix}
\mathds{I}\textbf{P}_1^- & \mathds{I}\textbf{P}_2^-\\[0.3em]
\end{bmatrix}\nonumber\\
&=\begin{bmatrix}
\textbf{P}_1^- & \textbf{P}_2^- \\[0.3em]
\textbf{P}_3^- & \textbf{P}_4^- \\[0.3em]
\end{bmatrix} - \begin{bmatrix}
\mathds{K}_1\textbf{L}^\intercal_1 & \mathds{K}_1\textbf{L}^\intercal_2\\[0.3em]
\mathds{K}_2\textbf{L}^\intercal_1 & \mathds{K}_2\textbf{L}^\intercal_2\\[0.3em]
\end{bmatrix}.
\end{align}

Specifically, measurement updates are given by
\begin{align}
\textbf{s}^+ &= \textbf{s}^- + \mathds{K}_1 \textbf{e}\label{Eq:SPlus}\\
\bm{\Omega}^+ &= \bm{\Omega}^- + \mathds{K}_2 \textbf{e}\label{Eq:OmegaPlus}\\
\textbf{P}^{+}_1 &= \textbf{P}^{-}_1 - \mathds{K}_1 \textbf{L}_1^\intercal\label{Eq:P1Plus}\\
\textbf{P}^{+}_2 &= \textbf{P}^{-}_2 - \mathds{K}_1 \textbf{L}_2^\intercal\label{Eq:P2Plus}\\
\textbf{P}^{+}_4 &= \textbf{P}^{-}_4 - \mathds{K}_2 \textbf{L}_2^\intercal\label{Eq:P4Plus}
\end{align}
The covariance blocks are initialized as follows
\begin{align}
\textbf{P}_1(0) &= \sigma^2_s \textbf{I}_{n_s}\\
\textbf{P}_4(0) &= \sigma^2_\Omega \textbf{I}_{n_{\bm{\Omega}}}\label{Eq:P4Init}\\
\textbf{P}_2(0) &= \textbf{0}\label{eq:P2Initial}.
\end{align}

The functional Bayesian filtering algorithm is summarized in Algorithm \ref{alg:FBF}.
\begin{algorithm}
\caption{Functional Bayesian Filter}\label{alg:FBF}
\begin{algorithmic}[h]
\State \textbf{Initialization:}
\State $n_u$: input dimension
\State $n_s$: state dimension
\State $a_s$: state kernel parameter
\State $a_u$: input kernel parameter
\State $\eta$: learning rate
\State $\sigma^2_s$: state variance
\State $\sigma^2_\Omega$: weight variance
\State $\sigma^2_y$: output variance
\State $\textbf{P}_1(0) = \sigma^2_s \textbf{I}_{n_s}$
\State $\textbf{P}_4(0) = \sigma^2_\Omega\textbf{I}_{n_{\bm{\Omega}}}$
\State $\textbf{P}_2(0) = \textbf{0}$
\State Randomly initialize input $\textbf{u}_0\in\mathbb{R}^{1\times n_u}$
\State Randomly initialize states $\textbf{s}_{-1}$ and $\textbf{s}_0\in\mathbb{R}^{1\times n_s}$
\State Randomly initialize coefficient matrix $\textbf{A}\in\mathbb{R}^{1\times n_s}$
\State ${\bm\Psi} = [\psi(\textbf{s}_{-1},\textbf{u}_0)]$: initial feature matrix
\State $\textbf{S} = [\textbf{s}_{-1}]$: initial state dictionary
\State $m =  1$: dictionary size
\State $\mathds{I}=\begin{bmatrix}
\textbf{0} & \textbf{I}_{n_y} \\[0.3em]
\end{bmatrix}\in\mathbb{R}^{n_y\times n_s}$: measurement matrix\\
\For{$i = 1,\cdots$}
\State 		\textbf{\underline{Predict}:}
\State 		Get current input $\textbf{u}_i$ and past state $\textbf{s}_i$
\State 		Propagate \textit{a priori} state estimate
\State 		$\textbf{s}_i={\bm\Omega}^\intercal_i\psi(\textbf{s}_{i-1},\textbf{u}_{i})$\hfill(\ref{algGenerateState})
\State 		Compute state transition dynamics
\State $\textbf{F}_1(i) = \underbrace{2a_s\textbf{A}^\intercal_i \textbf{K}_i\textbf{D}^\intercal_i}_{{\bm\varLambda}_i}$\hfill(\ref{Eq:F1})
\State 		$\textbf{F}_2(i) =\textbf{1}_{n_s}\psi(\textbf{s}_{i-1},\textbf{u}_{i})^\intercal $\hfill(\ref{Eq:F2})
\State 		Propagate \textit{a priori} covariance estimate\\
\State 		$\textbf{P}^{-}_1 = \left[\textbf{F}_1\textbf{P}^+_1 + \textbf{F}_2(\textbf{P}^+_2)^\intercal\right]\textbf{F}^\intercal_1$
\State $\quad \qquad + \underbrace{\left[\textbf{F}_1\textbf{P}^{+}_2+\textbf{F}_2\textbf{P}^{+}_4\right]}_{\textbf{P}_2^-} \textbf{F}_2^\intercal + \sigma^2_s\textbf{I}_{n_s}$\hfill(\ref{Eq:P1Min})
\State 	$\textbf{P}^{-}_4 = \textbf{P}^{+}_4+\sigma^2_\Omega \textbf{I}_{n_\psi}$\hfill(\ref{Eq:P4Min})
\State 	\textbf{\underline{Update}:}
\State 		Innovation or measurement residual
\State 		$\textbf{e}_i = \textbf{d}_i - \textbf{y}_i $
\State 		Compute Kalman gain
\State $\textbf{L}_1 \leftarrow$ last $n_y$ columns of $\textbf{P}_1^-$\hfill(\ref{Eq:L1})
\State 		$\textbf{L}_2 \leftarrow$ last $n_y$ columns of $(\textbf{P}_2^-)^\intercal$\hfill(\ref{Eq:L2})
\State 		$\textbf{M}\leftarrow$ the $n_y\times n_y$ lower-right corner of $\textbf{P}_1^-$\hfill(\ref{Eq:M})
\State   	$\textbf{N}\leftarrow (\textbf{M}+\sigma^2_y \textbf{I}_{n_y})^{-1}$\hfill(\ref{Eq:N})
\State 			$\mathds{K}_1\leftarrow \textbf{L}_1 \textbf{N}$\hfill(\ref{Eq:K})
\State 		$\mathds{K}_2\leftarrow \textbf{L}_2 \textbf{N}$\hfill(\ref{Eq:K})
\State 		\textbf{\underline{Update \textit{a posteriori} estimates}:}	
\State 		$\textbf{s}^+ = \textbf{s}^- + \mathds{K}_1 \textbf{e}$\hfill(\ref{Eq:SPlus})
\State 		$\bm{\Omega}^+ = \bm{\Omega}^- + \mathds{K}_2 \textbf{e}$\hfill(\ref{Eq:OmegaPlus})
\State 		$\textbf{P}^{+}_1 = \textbf{P}^{-}_1 - \mathds{K}_1 \textbf{L}_1^\intercal$\hfill(\ref{Eq:P1Plus})
\State 		$\textbf{P}^{+}_2 = \textbf{P}^{-}_2 - \mathds{K}_1 \textbf{L}_2^\intercal$\hfill(\ref{Eq:P2Plus})
\State 			$\textbf{P}^{+}_4 = \textbf{P}^{-}_4 - \mathds{K}_2 \textbf{L}_2^\intercal$\hfill(\ref{Eq:P4Plus})
\EndFor
\end{algorithmic}
\end{algorithm}

Functional Bayesian filtering requires the state-space model be expressed as a linear model in the RKHS (using the representer theorem, as a finite linear combination of kernel products evaluated on the training data). This can be learned directly from observations (from either known or unknown system dynamics). Once the generative model, in terms of the kernel filter weights $\bm{\Omega}$, is obtained or fixed (after training), Algorithm \ref{alg:FBF} reduces to just the state update, with state $\textbf{x} = \textbf{s}_i$ in (\ref{Eq:AugState}), state transition equation $\textbf{F} = \textbf{F}_1$ in (\ref{eq:StateTrans}), and covariance $\textbf{P} = \textbf{P}_1$ in (\ref{eq:P}).
\subsection{Computation in input space using kernel trick}
Here, we show how to compute each of the submatrices $\textbf{P}^-_i$  of the \textit{a priori} state covariance matrix estimate $\textbf{P}$ in (\ref{Eq:P1Min}-\ref{Eq:P4Min}), except $\textbf{P}^-_3$, since it is simply the transpose of $\textbf{P}^-_2$. In order to make the actual computation tractable, we break the weight matrix vector $\bm{\Omega}_i$ in (\ref{Eq:AugState}) into individual state dimension components and rewrite (\ref{Eq:SSM}) as
\begin{align}
\begin{bmatrix}
\textbf{s}_i \\[0.3em]
\bm{\Omega}^{(k)}_i\\[0.3em]
\end{bmatrix} = \begin{bmatrix}
\textbf{F}_1 & \textbf{F}^{(k)}_2 \\[0.3em]
\textbf{0} & \textbf{I}_{n_{\Omega^{(k)}}} \\[0.3em]
\end{bmatrix}\begin{bmatrix}
\textbf{s}_{i-1} \\[0.3em]
\bm{\Omega}^{(k)}_i\\[0.3em]
\end{bmatrix} + \textbf{w}_{i-1} \label{Eq:SSM_k}
\end{align}
where $\bm{\Omega}^{(k)}_i$ are all the weights connected to the $k$-th output state node, with $1\leq k\leq n_s$, and
\begin{align}
\textbf{F}_2(i)^{(k)} \stackrel{\Delta}{=}\frac{\partial \textbf{s}_i}{\partial \bm{\Omega}^{(k)}_i} = \textbf{I}^{(k)}_{n_s}\psi(\textbf{s}_{i-1},\textbf{u}_{i})^\intercal \label{Eq:F2new}.
\end{align}
At each time step $i$, this process is repeated for each of the $n_s$ state components. 
\subsubsection{Computing $\textbf{P}^-_2$ Recursively}
First, we show how to recursively update the weights covariance matrix $(\textbf{P}^-_2)^{(k)}\in\mathbb{R}^{n_s\times n_{\bm{\Omega}^{(k)}}}$, where $n_\psi$ is infinite for the Gaussian kernel.
Substituting (\ref{Eq:F1}), (\ref{Eq:F2new}), and (\ref{Eq:P2Plus}) into (\ref{eq:P2}), for the $k$th state component, gives
\begin{align}
\left(\textbf{P}^{-}_2\right)^{(k)} &= {\bm\varLambda}_i\left(\textbf{P}^{+}_2\right)^{(k)}+\textbf{I}^{(k)}_{n_s}\psi(\textbf{s}_{i-1},\textbf{u}_{i})^\intercal\left(\textbf{P}^{+}_4\right)^{(k)}\nonumber\\
&=\scalebox{0.85}{${\bm\varLambda}_i\left(\textbf{P}^{-}_2 - \mathds{K}_1 \textbf{L}_2^\intercal\right)^{(k)}+\textbf{I}^{(k)}_{n_s}\psi(\textbf{s}_{i-1},\textbf{u}_{i})^\intercal\left(\textbf{P}^{+}_4\right)^{(k)}$}\nonumber\\
&=\scalebox{.85}{${\bm\varLambda}_i\left(\textbf{P}^{-}_2 - \mathds{K}_1 \mathds{I}\textbf{P}^{-}_2\right)^{(k)}+\textbf{I}^{(k)}_{n_s}\psi(\textbf{s}_{i-1},\textbf{u}_{i})^\intercal\left(\textbf{P}^{+}_4\right)^{(k)}$}\nonumber\\
&=\scalebox{.85}{$\underbrace{{\bm\varLambda}_i\left(\textbf{I}_{n_s} - \mathds{K}_1 \mathds{I}\right)}_{{\bm\varLambda}'_i}\left(\textbf{P}^{-}_2\right)^{(k)}+\textbf{I}^{(k)}_{n_s}\psi(\textbf{s}_{i-1},\textbf{u}_{i})^\intercal\left(\textbf{P}^{+}_4\right)^{(k)}$}\nonumber\\
&\stackrel{\Delta}{=}{\bm\varLambda}'_i\left(\textbf{P}^{-}_2\right)^{(k)}+\textbf{I}^{(k)}_{n_s}\psi(\textbf{s}_{i-1},\textbf{u}_{i})^\intercal\left(\textbf{P}^{+}_4\right)^{(k)}.\label{eq:P2_new}
\end{align}
The above expression has the same form as Equation (28) in the kernel adaptive ARMA formulation \cite{KAARMA}, which we reproduce here:
\begin{align}
\frac{\partial \textbf{s}_i}{\partial {\bm\Omega}_i^{(k)}} &= {\bm\varLambda}_i\frac{\partial\textbf{s}_{i-1}}{\partial{\bm\Omega}_i^{(k)}} + \textbf{I}^{(k)}_{n_s}\psi(\textbf{s}_{i-1},\textbf{u}_{i})^\intercal.\label{eq:GradientDescent}
\end{align} 
We see that the FBF formulation reduces to gradient descent if the state variable Kalman gain $\mathds{K}_1$ is zero and the weights covariance matrix $\left(\textbf{P}^{+}_4\right)^{(k)}$ is the identity matrix, i.e., unit variance. In other words, the FBF generalizes the kernel adaptive ARMA algorithm. Using the information theoretic interpretation, where the state-gradient gradient ${\bm\varLambda}_i$ is a weighted Parzen estimate of the joint input data $(\textbf{s}_i,\textbf{u}_i)$ pdf, we are recursively evolving the density.

Given the initialization in (\ref{eq:P2Initial}), the ensuing update becomes
\begin{align}
\textbf{P}^{(k)}_2 (1|0) &=\textbf{I}^{(k)}_{n_s}\psi(\textbf{s}_{0},\textbf{u}_{1})^\intercal\textbf{P}^{(k)}_4(0).
\end{align}
By induction, we can factor out the basis functions $\psi(\cdot,\cdot)$ and express the recursion (\ref{eq:P2_new}) as
\begin{align}
\left(\textbf{P}^{-}_2\right)^{(k)} &= {\bm\varLambda}'_i\textit{\textbf{V}}^{(k)}_{i-1} {\bm\Psi'}^\intercal_{i-1} + \textbf{I}^{(k)}_{n_s}\psi(\textbf{s}_{i-1},\textbf{u}_{i})^\intercal\left(\textbf{P}^{+}_4\right)^{(k)}\nonumber\\
&=\left[{\bm\varLambda}'_i \textbf{\textit{V}}^{(k)}_{i-1} , \textbf{I}^{(k)}_{n_s}\right]\begin{bmatrix}
{\bm\Psi'}_{i-1}^\intercal \\[0.3em]
\psi(\textbf{s}_{i-1},\textbf{u}_{i})^\intercal\left(\textbf{P}^{+}_4\right)^{(k)} \\[0.3em]
\end{bmatrix}\\
&=\textit{\textbf{V}}^{(k)}_{i} {\bm\Psi'}^\intercal_{i}\label{Eq:P2Update}
\end{align}
where ${\bm\Psi'}_{i} \stackrel{\Delta}{=} [{\bm\Psi'}_{i-1}, \left(\textbf{P}^{+}_4\right)^{(k)}\psi(\textbf{s}_{i-1},\textbf{u}_{i})]\in\mathbb{R}^{n_{\psi}\times i}$ are centers generated by the input sequence and forward-propagated states, normalized by $\textbf{P}^{(k)}_4(i-1|i-1)$, and $\textit{\textbf{V}}^{(k)}_{i}\stackrel{\Delta}{=} \left[{\bm\varLambda}'_i \textit{\textbf{V}}^{(k)}_{i-1} , \textbf{I}^{(k)}_{n_s}\right]\in\mathbb{R}^{n_s\times i}$ is the updated state-transition gradient, with initializations ${\bm\Psi'}_1 = [\sigma^2_{\Omega}\psi(\textbf{s}_{0},\textbf{u}_1) ]$ and $\textit{\textbf{V}}^{(k)}_1 =  \textbf{I}^{(k)}_{n_s}$.

\subsubsection{Updating $\textbf{P}_4$}
Here we show that $\textbf{P}_4$ maintains its diagonal form after the measurement update, given its initialization in (\ref{Eq:P4Init}):
\begin{align}
\left(\textbf{P}^{+}_4\right)^{(k)} &= \left(\textbf{P}^{-}_4\right)^{(k)} - \left(\mathds{K}_2 \textbf{L}_2^\intercal\right)^{(k)}\\
&= \scalebox{.9}{$\left(\sigma^{2-}_\Omega\right)^{(k)} \textbf{I}_{n_{\bm{\Omega}}} - \left(\textbf{P}^{-\intercal}_2\mathds{I}^\intercal(\textbf{M}+\sigma^2_y \textbf{I}_{n_y})^{-1}\mathds{I}\textbf{P}^{-}_2\right)^{(k)}$}\\
&= \left(\sigma^{2-}_\Omega\right)^{(k)}  \textbf{I}_{n_{\bm{\Omega}}}- {\bm\Psi'}\underbrace{\textit{\textbf{V}}^{(k)\intercal}_{i}\mathds{I}^\intercal\textbf{N}\mathds{I}\textit{\textbf{V}}^{(k)}_{i}}_{\textbf{B}_i^{(k)}} {\bm\Psi'}^\intercal_{i}\\
&= \scalebox{.6}{$\left(\sigma^{2-}_\Omega\right)^{(k)}  \textbf{I}_{n_{\bm{\Omega}}} - \underbrace{[{\bm\Psi'}_{i-1}, \left(\sigma^{2}_\Omega\right)^{(k)}\psi(\textbf{s}_{i-1},\textbf{u}_{i})]\textbf{B}_i^{(k)} \begin{bmatrix}
	{\bm\Psi'}^\intercal_{i-1} \\[0.3em]
	\left(\sigma^{2}_\Omega\right)^{(k)}\psi(\textbf{s}_{i-1},\textbf{u}_{i}) \\[0.3em]
	\end{bmatrix}}_{\left(\varsigma^2_i\right)^{(k)} }$}\label{Eq:varsigma}\\
&= \underbrace{\left((\sigma^{2-}_\Omega)^{(k)} - (\varsigma^2_i)^{(k)}\right)}_{(\sigma^{2+}_\Omega)^{(k)}}\textbf{I}_{n_{\bm{\Omega}}}\label{Eq:P4Plus}
\end{align}
where $\textbf{B}_i^{(k)}\stackrel{\Delta}{=}\textit{\textbf{V}}^{(k)\intercal}_{i}\mathds{I}^\intercal\textbf{N}\mathds{I}\textit{\textbf{V}}^{(k)}_{i}\in\mathbb{R}^{i\times i}$ can be computed in a straightforward manner, and if we denote its $i$th row and $j$th column element as $b_{ij}^{(k)}$, then the entire second term on the right-hand side of (\ref{Eq:varsigma}) becomes a scalar, using the kernel trick:
\begin{align}
(\varsigma^2_i)^{(k)} &= \scalebox{.95}{$\sum_j\sum_i (\sigma^{2}_\Omega)^{(k)}_i\psi(\textbf{s}_{i-1},\textbf{u}_{i})b_{ij}^{(k)}(\sigma^{2}_\Omega)^{(k)}_j\psi(\textbf{s}_{j-1},\textbf{u}_{j})$}\nonumber\\
&= \sum_j\sum_i b_{ij}^{(k)}(\sigma^{2}_\Omega)^{(k)}_i(\sigma^{2}_\Omega)^{(k)}_j\mathcal{K}(\textbf{s}_{i},\textbf{s}_{j})\mathcal{K}(\textbf{u}_{i},\textbf{u}_{j}).
\end{align}
Similarly, we see that there is an information theoretic interpretation, where the RKHS weight covariance $\textbf{P}_4$ is updated using a weighted information potential of the joint input data  $(\textbf{s}_i,\textbf{u}_i)$.

From (\ref{Eq:P4Plus}), we can simply substitute $(\textbf{P}^+_4)^{(k)}$ with the scalar weighted identity matrix $(\sigma^{2+}_\Omega)^{(k)}\textbf{I}_{n_{\bm{\Omega}}}$, which we denote the scalar in shorthand as $\rho_{i-1} \stackrel{\Delta}{=}\left(\sigma^{2+}_\Omega\right)^{(k)}(i-1|i-1)$. Specifically, the weighted features in (\ref{Eq:P2Update}) simplifies to 
\begin{align}
{\bm\Psi'}_{i} \stackrel{\Delta}{=} [{\bm\Psi'}_{i-1}, \psi(\textbf{s}_{i-1},\textbf{u}_{i})]\label{Eq:newWF}
\end{align}and
\begin{align}
\textit{\textbf{V}}^{(k)}_{i}\stackrel{\Delta}{=} \left[{\bm\varLambda}'_i \textit{\textbf{V}}^{(k)}_{i-1}, \rho_{i-1}\textbf{I}^{(k)}_{n_s}\right]\label{Eq:newC}
\end{align}
where we can now separate the feature centers completely from their coefficients for ease of computation.

\subsubsection{Updating $\textbf{P}^-_1$}
Given the derivations in (\ref{Eq:P4Plus}), (\ref{Eq:newWF}), and (\ref{Eq:newC}), computing the state variable covariance matrix $\textbf{P}^-_1$ becomes straightforward. In order to unclutter the notation, we first show some intermediate steps. Substituting (\ref{Eq:F2new}) and (\ref{Eq:P2Plus}) for $\textbf{F}_2^{(k)}$ and $(\textbf{P}^+_2)^{(k)}$, respectively, the following product can be expressed as
\begin{align}
\textbf{F}_2^{(k)}((\textbf{P}^+_2)^{(k)})^\intercal &= \scalebox{.85}{$\textbf{I}_{n_s}^{(k)}\psi(\textbf{s}_{i-1},\textbf{u}_{i})^\intercal\left(\left(\textbf{I}_{n_s} - \mathds{K}_1 \mathds{I}\right)\left(\textbf{P}^{-}_2\right)^{(k)}\right)^\intercal$}\nonumber\\
&\stackrel{(a)}{=}\scalebox{.85}{$\textbf{I}_{n_s}^{(k)}\underbrace{\psi(\textbf{s}_{i-1},\textbf{u}_{i})^\intercal{\bm\Psi'}_{i-1}}_{\textbf{k}_{i-1}}\textit{\textbf{V}}^{(k)\intercal}_{i-1}\left(\textbf{I}_{n_s} - \mathds{K}_1 \mathds{I}\right)^\intercal$}
\end{align}
where equality $(a)$ follows from (\ref{Eq:P2Update}), and $\textbf{k}_{i-1}$ is a vector of kernel evaluations:
\begin{align}
\textbf{k}_{i-1} = \begin{bmatrix}
\mathcal{K}(\textbf{s}_{i-1},\textbf{s}_{0})\mathcal{K}(\textbf{u}_{i},\textbf{u}_{1}) \\[0.3em]
\vdots \\[0.3em]
\mathcal{K}(\textbf{s}_{i-1},\textbf{s}_{i-2})\mathcal{K}(\textbf{u}_{i},\textbf{u}_{i-1}) \\[0.3em]
\end{bmatrix}.\label{Eq:FP}
\end{align}
Similarly
\begin{align}
(\textbf{P}^-_2)^{(k)}(\textbf{F}_2^{(k)})^\intercal &=
\textit{\textbf{V}}^{(k)}_{i}\underbrace{{\bm\Psi'}^\intercal_{i}\psi(\textbf{s}_{i-1},\textbf{u}_{i})}_{\textbf{k}^\intercal_{i}}(\textbf{I}_{n_s}^{(k)})^\intercal.\label{Eq:PF}
\end{align}
Finally, rewriting the \textit{a priori} state covariance estimate in (\ref{Eq:P1Min}) for the $k$-th state component, using (\ref{Eq:FP}) and (\ref{Eq:PF}), yields
\begin{align}
(\textbf{P}^{-}_1)^{(k)} &= \left[\textbf{F}_1(\textbf{P}^+_1)^{(k)} + (\textbf{F}_2)^{(k)}((\textbf{P}^+_2)^{(k)})^\intercal\right]\textbf{F}^\intercal_1 \nonumber\\
&\qquad + (\textbf{P}_2^-)^{(k)}(\textbf{F}_2^{(k)})^\intercal+\sigma^2_U \textbf{I}^{(k)}_{n_s}\nonumber\\
&= \scalebox{.95}{$\left[{\bm\varLambda}_i(\textbf{P}^+_1)^{(k)} + \textbf{I}_{n_s}^{(k)}\textbf{k}_{i-1}\textit{\textbf{V}}^{(k)\intercal}_{i-1}\left(\textbf{I}_{n_s} - \mathds{K}_1 \mathds{I}\right)^\intercal\right]{\bm\varLambda}_i^\intercal$} \nonumber\\
&\qquad + \scalebox{.95}{$ \textit{\textbf{V}}^{(k)}_{i}\textbf{k}^\intercal_{i}(\textbf{I}_{n_s}^{(k)})^\intercal + \sigma^2_U \textbf{I}^{(k)}_{n_s}$}.
\end{align}
which can be calculated in a straightforward manner, since all the terms involved are now expressed using finite dimensional vectors and matrices.
\subsection{Complexity}
The FBF memory and computational complexities for each recursive update is $O(n)$ and $O(n^2)$. Unlike gradient descent learning in KAARMA \cite{KAARMA}, only a single state vector is produced per update, in the current trajectory.
\subsection{The Kernel Advantage}
The major contribution of this paper is  a novel formulation under a unifying framework that tackles all three major shortcomings of classic Kalman filter: linearity, prior knowledge of accurate system parameters, and Gaussian assumption. 

Kernel methods have additional properties that make them especially appealing for real-world applications. For example, we can exploit the structure of the RKHS to our advantage as we already demonstrated with the nearest instance centroid estimation (NICE) approach [20]. In fact, the representer theorem yields an input-output function that is a sum of terms centered at the data samples. We can partition this sum into quasi-orthogonal sub-sums, simplifying the processing and allowing the design of novel input-output maps by simple concatenation, which opens the door for fast search procedure to compose on-the-fly new filters using ideas from transfer learning, as required for nonstationary environments. For Bayesian filtering, it is crucial that the two basic properties of an error covariance are preserved in each iteration, namely symmetry and positive definiteness. In practice, due to numerical errors by arithmetic operations, these two properties are often lost or destroyed and a square-root formulation is required \cite{Arasaratnam2009}. Kernel methods, on the other hand, are immune to these problems due to the use of positive definite kernel function satisfying Mercer's conditions.

Furthermore, kernel methods have the added advantage of preserves the learning algorithm regardless of input type (numerical or nonnumerical), by selecting the appropriate reproducing kernel. This formulation imposes no restriction on the relationship between the input signals. This is important for input signals having different representations and spatiotemporal scales, e.g., we can model a biological system, taking spike trains, continuous amplitude local field potentials (LFPs), and vectorized state variables as inputs.
\section{Experiments and Results}
Here, we illustrate and evaluate the proposed Functional Bayesian Filter using numerical examples. As a proof-of-concept, we consider the following tasks: recurrent network training, chaotic time-series estimation and cooperative filtering using Gaussian and non-Gaussian noises, and inverse kinematics modeling.

\subsection{Cooperative Filtering for Signal Enhancement}
First, we consider the scenario of an unknown nonlinear system with only noisy observations available, and compare the functional Bayesian filter with cubature Bayesian filter in training a dynamical system in the form of recurrent network on the Mackey-Glass (MG) chaotic time series \cite{Mackey77}, defined by the following time-delay ordinary differential equation
\begin{align*}
\frac{d x(t)}{d t} =\frac{\beta x(t-\tau)}{1+x(t-\tau)^{n}} -\gamma x(t)
\end{align*}
where $\beta=0.2$, $\gamma=0.1$, $\tau=30$, $n=10$, discretized at a sampling period of 6 seconds using the forth-order Runge-Kutta method, with initial condition $x(t) = 0.9$. Chaotic dynamics are extremely sensitive to initial conditions: small differences in initial conditions yields widely diverging outcomes, rendering long-term prediction intractable, in general. 

\begin{figure}[h]
	\centering
	\includegraphics[width=0.4\textwidth]{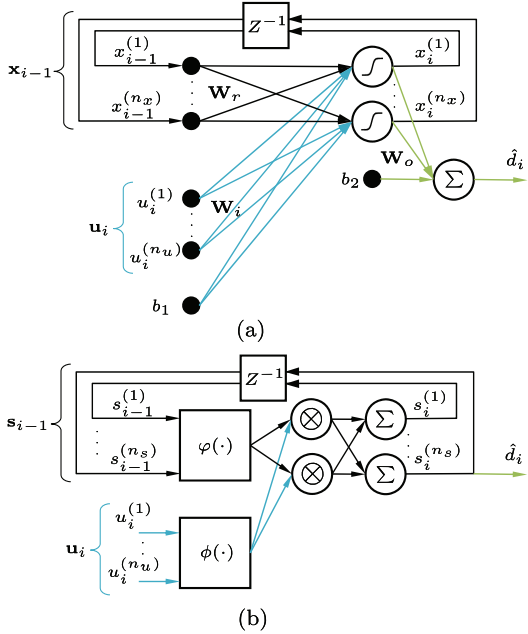}
	\caption{Recurrent network trained using (a) Square-Root Cubature Kalman Filter (b) Functional Bayesian Filter.}
	\label{fig:RNN_arch}
\end{figure}

Cooperative filtering aims to construct an empirical model using (pseudo-) clean data extracted from the noisy measurements. The signal estimator is coupled with the weight parameter estimator. Here, we following the experimental setup outlined in \cite{Arasaratnam2009}. For the cubature Kalman approach (specifically, the square-root version or SCKF), an RNN is used to model the dynamics with its weights, the weights of the RNN are estimated from the latest signal estimate (and \textit{vice versa}).   The state-space model for the RNN architecture, trained using square-root cubature Kalman filter (SCKF), is defined as
\begin{align*}
\textbf{w}_k &= \textbf{w}_{k+1} + \textbf{q}_{k-1}\\
d_k &= W_o {\rm h}h(W_r\textbf{x}_{k-1}+W_i\textbf{u}_k) +r_k 
\end{align*}
where $W_i,W_r,$ and $W_o$ are the input, recurrent, and output weight matrices of appropriate dimensions (collectively, they form the weight vector $\textbf{w}_k$), the process noise is additive Gaussian with zero-mean and covariance $Q_{k-1}$, i.e., $\textbf{q}_k\sim \mathcal{N}(0,Q_{k-1})$, the measurement noise is $\textbf{r}_k\sim \mathcal{N}(0,R_{k})$, internal state or output of the hidden layer at time $(k-1)$ is $\textbf{x}_{k-1}$, the input is denoted $\textbf{u}_k$, the desired output $d_K$ is the measurement, and ${\rm h}h(\cdot)$ denotes the activation function. The input embedding dimension is set at $n_u = 7$, with one self-recurrent hidden layer with 5 neurons ($n_s = 5$), a single output, and bias at each node, as shown in Fig. \ref{fig:RNN_arch}(a). The hidden neuron activations use hyperbolic tangent function, the output neuron is linear.
The dynamic state-space model for the signal estimator is written as
\begin{align*}
\textbf{u}_k &= \textbf{f}(\textbf{u}_{k-1},\hat{\textbf{w}}_{k-1|k-1},\textbf{x}_{k-2}) + [ 1\quad 0\cdots 0]^\intercal v_{k+1}\\
z_k &= [1\quad 0\cdots 0]\textbf{u}_k  +e_k 
\end{align*}
where $\textbf{u}_k =[u_k\quad u_{k-1}\cdots u_{k-6}]$ is the data window, $\textbf{f}$ denotes the 7-5R-1 RNN state transition function concatenated with the delayed output, the measurement noise $e_k\sim\mathcal{N}(0,\sigma^2_e)$ corresponds to the SNR, the process noise $v_{k-1}\sim\mathcal{N}(0,\sigma^2_v)$ was fixed to be $10\%$ of $\sigma^2_e$), and the initial estimate is assumed to be zero with unity covariance.

A noisy (signal-to-noise ratio fixed at 10dB) MG chaotic time sequence of 1000 samples are used to train the RNN. For each training run, ten batches were made. Each batch consists of 100 time-step updates, from a randomly selected starting point in the training sequence. The state of the RNN at $t=0$ was assumed to be zero, i.e., $\textbf{x}_0=\textbf{0}$. During the test phase, SCKF is performed on an independent sequence of 100 noisy samples using the state-transition equation obtained from the fixed weights $\textbf{w}$. The ensemble-averaged mean square error (MSE) is computed for 50 independent training runs. 

For FBF, the recurrent architecture is parsimonious (Fig. \ref{fig:RNN_arch}(b)), and the signal state and network weights are estimated simultaneous by construction. The kernel parameters for the state, input, state covariance P1, and weigth covariance P4 are $a_s = 0.6$, $a_u=1.8$, $a_{P_1}= 0.4$, and $a_{P_4} = 0.2$, respectively. The state covariance, output variance, and weight covariance are initialized as $\sigma^2_{s} = 10$, $\sigma^2_{y} = 0.08$, $\sigma^2_{P_4} = 10$, respectively. There are no bias terms for our FBF implementation. Using the small-step-size theory framework \cite{Haykin02}, which self-regularizes kernel adaptive filter, we scale the state Kalman gain $\mathds{K}_1$ by a constant factor of 0.5, and the weight gain $\mathds{K}_1$ by a constant factor of 0.1.

\begin{figure}[h]
	\centering
	\includegraphics[width=0.458\textwidth]{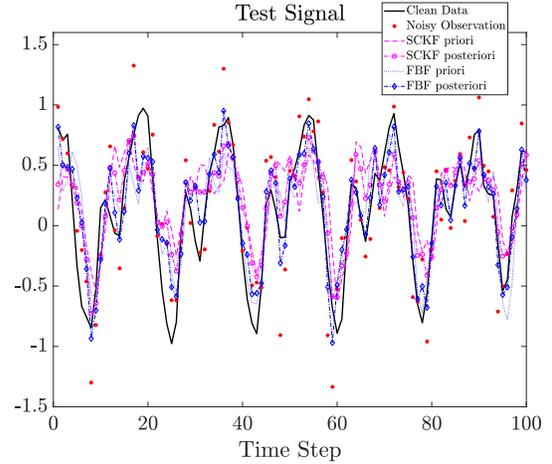}
	\caption{Test signal filtering.}
	\label{fig:MGplot}
\end{figure}

\begin{figure}[t!]
	\centering
		\begin{subfigure}
		\centering
		\includegraphics[width=0.46\textwidth]{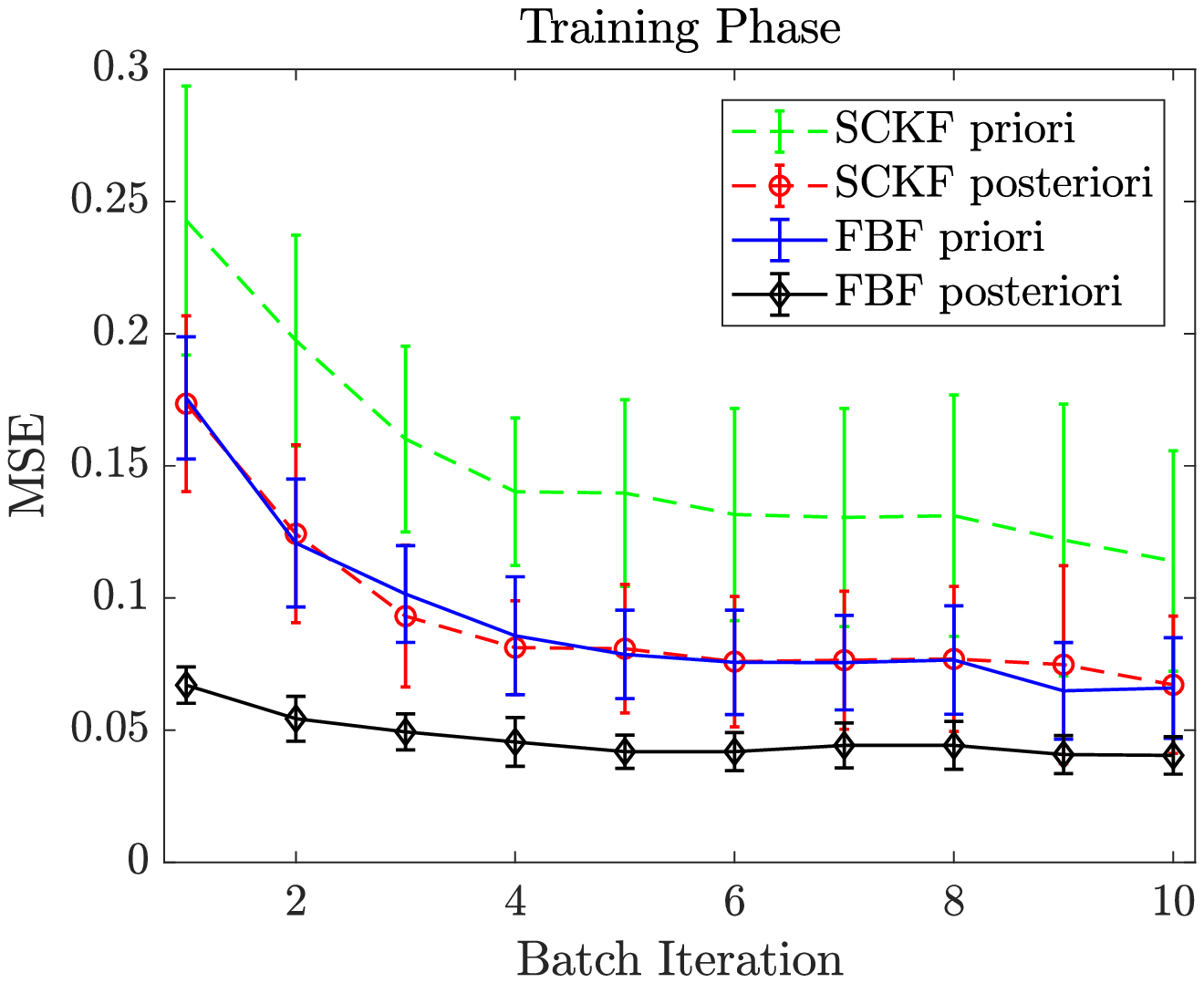}
	\end{subfigure}%
	\begin{subfigure}
		\centering
		\includegraphics[width=0.46\textwidth]{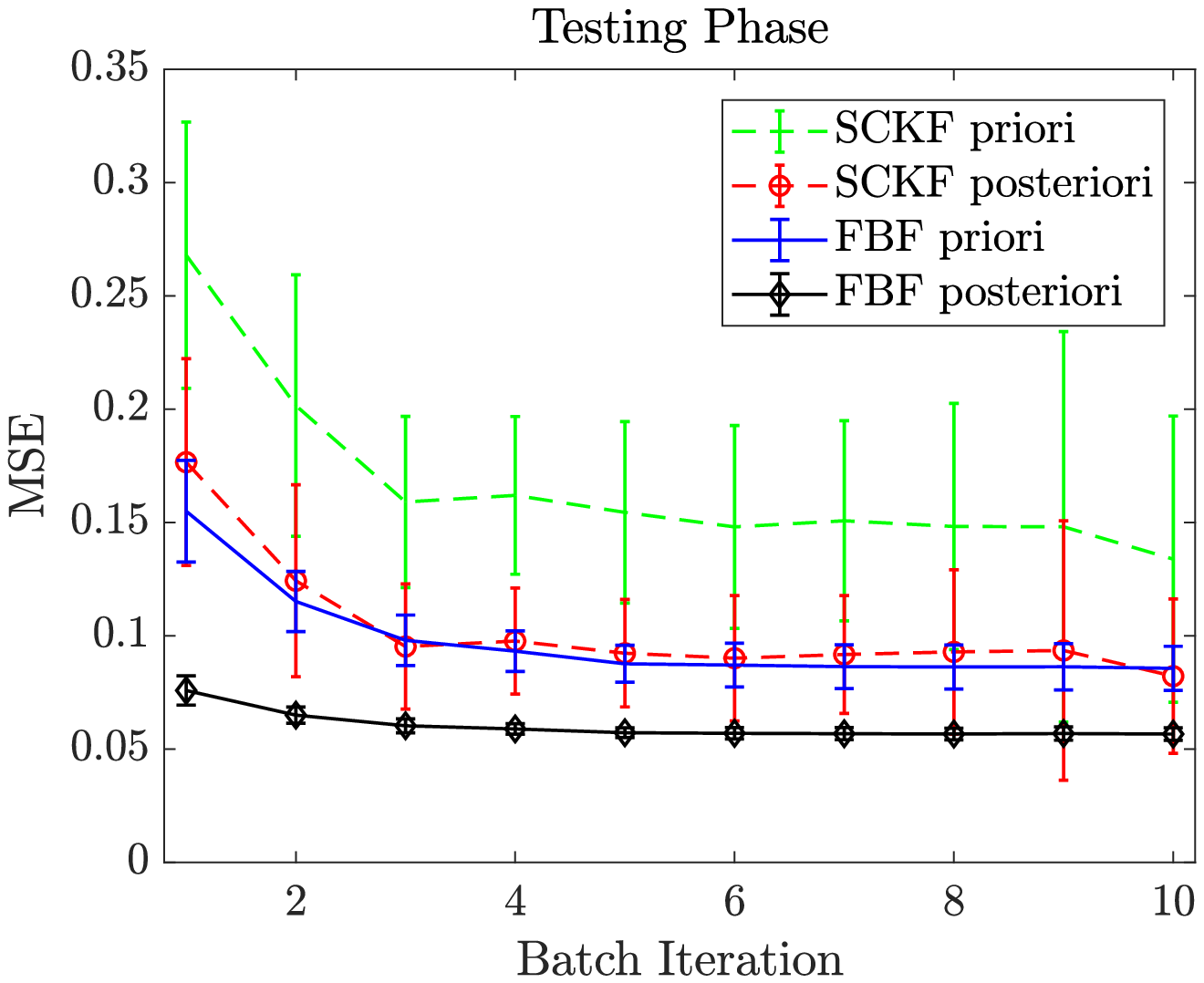}
	\end{subfigure}
		\caption{Ensemble-averaged Mean-Squared Error (MSE) over 50 runs vs. number of batch iterations (each training iteration consists of a 100-sample sequence with random starting point).}
	\label{fig:MGperformance}
\end{figure}

SCKF has been successfully validated to significantly outperform other known nonlinear filters such as EKF and central-difference Kalman filter (CDKF) and provides improved numerical stability over CKF \cite{Arasaratnam2009}. It's important to note that the two properties of error covariance matrix (symmetry and positive definiteness) are always preserved in FBF, since we are using positive definite kernel function satisfying Mercer's conditions, unlike input-space arithmetic operations such as CKF, where these two properties are often lost or destroyed and a square-root version is preferred. Here we focus on the performance comparison of SCKF and FBF. Fig. \ref{fig:MGplot} shows the filtering performance on the independent test signal during one of the 50 runs. The ``priori'' label denotes the time update using the predictive density before receiving a new measurement; ``posteriori'', the measurement update from the posterior density.  Fig. \ref{fig:MGperformance} shows the ensemble-averaged MSE (error bars represent one standard deviation) over all 50 runs versus the number of batch iterations, where each training iteration consists of a 100-sample noisy sequence with random starting point in the 1000-sample training data.

Clearly, the FBF significantly improves the quality of the signal as compared to the SCKF.

\subsection{Ikeda}
Next, we evaluate the performance of the FBF using multivariate chaotic time series and under various non-Gaussian noise conditions. The 2D example of the Ikeda map is defined by
\begin{align}
\scriptsize
f_{\rm Ikeda}(x_{i+1},y_{i+1})=\begin{cases}
1 + u (x_i \cos t_i - y_i \sin t_i)\\
u (x_i \sin t_i + y_i \cos t_i)
\normalsize
\end{cases}
\end{align}
where parameter $u=0.84$, $t_i = 0.4 - \frac{6}{1+x_n^2+y_n^2}$, and initial condition $[x_0,y_0]^\intercal = [1,0]^\intercal$. Four different types of additive noise (Gaussian, Laplacian, uniform, and alpha stable) is introduced to clean Ikeda data to obtain the noisy data $\{\textbf{y}_i\}$, with signal-to-noise ratio (SNR) of 3 dB, from which we will estimate the noiseless data. The initial 201 data points are used for training and the next 200 points are for testing. For each estimator, 20 independent trials were used to produce the MSE result (mean $\pm$ $\sigma$). The nonlinear Kalman extensions (EKF, UKF, and CKF) assume known accurate system models, which should provide an advantage. For DSMCE, the conditional embedding operator construction assume known hidden states $\{\textbf{x}_i\}$ and noisy measurements $\{\textbf{y}_i\}$, which should also provide an advantage during training. For KKF-CEO and FBF, only the noisy measurements are used for training. For a detailed discussion of the experimental set up, please refer to \cite{Zhu14CEO}.

For FBF, we trained using hidden states $\textbf{x}_i$ of dimension $n_x = 2$, i.e., the augmented state vectors $\textbf{s}_i$ have dimension $n_s = 4$, with input state kernel parameter $a_s = 0.8$, state covariance kernel parameter $a_{\textbf{P}_1} = 0.8$, and weight covariance kernel parameter $a_{\textbf{P}_4} = 0.8$. Initialization was set for the following parameters: state variance $\sigma^2_s = 1$, weight variance $\sigma^2_\Omega = 1$, output variance $\sigma^2_y = 40$.

\begin{table*}[ht]\renewcommand{\arraystretch}{1.5}
	\scriptsize
	\centering\caption{Test Set MSE of 2D Ikeda Map}
	\label{tab:SvN}
	\begin{tabular}{ |c|c|c|c|c|c|c|}
		\hline		
		\diagbox[width=5em]{Noise}{Alg.} & EKF & UKF & CKF & DSMCE & KKF-CEO & \textbf{FBF} \\ \hline
		Gaussian& $0.3630\pm 0.0448$ & $0.2639\pm 0.0218$ & $0.2374\pm 0.0176$ & $0.3918\pm 0.0502$ & $0.2253\pm 0.0168$ & \boldmath{$0.1803 \pm 0.0129$} \\ \hline
		Laplacian & $0.3897\pm0.0407$ & $0.2719\pm0.0217$ & $0.2574\pm0.0199$ & $0.3555\pm0.0626$ & $0.2121\pm0.0202$ & \boldmath{$0.1687 \pm 0.0117$} \\ \hline
		Uniform & $0.3843\pm0.0308$ & $0.2696\pm0.0213$ & $0.2427\pm0.0193$ & $0.3945\pm0.0800$ & $0.2384\pm0.0180$ & \boldmath{$0.1848\pm 0.0103$} \\ \hline
		Stable ({\tiny $\alpha = 1.6$}) & $0.3021\pm0.1232$ & $0.2465\pm0.0969$ & $0.2580\pm0.0991$ & $0.2319\pm0.1335$ & $0.1461\pm0.0413$ & \boldmath{$0.1224\pm 0.0205$}  \\ \hline
	\end{tabular}
	\normalsize
\end{table*}

Table \ref{tab:SvN} summarizes the mean-squared-error (MSE) testing set performance of the proposed FBF with the performances of EKF, UKF, CKF, DSMCE, and KKF-CEO. Despite the obvious disadvantage of not having accurate system knowledge or access to hidden states during training, FBF outperforms the best in all four noisy environments tested. The next-best performances are given by KKF-CEO. Kernel methods are able to leverage high-dimensional nonlinear representation of the signal in the RKHS to better model dynamical systems. Furthermore, the FBF uses informations theoretic learning to preserve the nonparametric nature of correlation learning and MSE adaptation. The cost function is still directly estimated from observation via a Parzen kernel estimator, but it extracts more infomration from the data for adaptation, and therefore yields solutions that are more accurate than MSE in non-Gaussian and nonlinear signal processing. The FBF outperformed the KKF-CEO method because it uses a full state-space representation constructed in the RKHS, which can scale with the complexity of the nonlinear dynamics, and not only assuming a simple additive noise system model.

\subsection{Modeling Inverse Kinematics in a Robotic Arm}
\begin{figure}[h]
	\centering
	\includegraphics[width=0.25\textwidth]{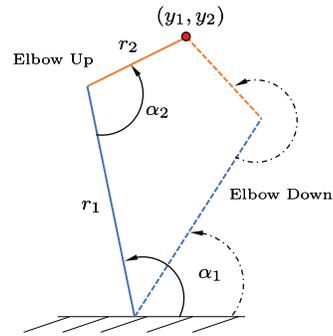}
	\caption{Two-joint robot arm illustrating how the Cartesian coordinates $(y_1, y_2)$ of the end effector is mapped to the given angles $(\alpha_1, \alpha_2)$. The solid and dashed lines show the `elbow up' and `elbow down' situations, respectively.}
	\label{fig:arm}
\end{figure}

In a two-joint robotic arm, Fig. \ref{fig:arm}, given the joint angles $(\alpha_1,\alpha_2)$, the kinematics equations give the Cartesian coordinate of the robot arm end-effector position as:
\begin{align*}
\scriptsize
(y_1,y_2)=\begin{cases}
r_1\cos(\alpha_1)-r_2\cos(\alpha_1+\alpha_2)\\
r_1\sin(\alpha_1)-r_2\sin(\alpha_1+\alpha_2)
\end{cases}
\normalsize
\end{align*}
where $r_1 = 0.8$ and $r_2 = 0.2$ are the link lengths, with $\alpha_1 \in [0.3,1.2]$ and $\alpha_2 \in [\pi/2,3\pi/2]$ are the joint ranges. Finding the mapping from $(y_1,y_2)$ to $(\alpha_1,\alpha_2)$ is called the \textit{inverse kinematics}, which is not a one-to-one mapping: as shown in Fig. \ref{fig:arm}, both the elbow-up and elbow-down joint angles result in the same tip-of-the-arm position.

Let the state vector be $\textbf{x}=[\alpha_1,\alpha_2]^\intercal$, and the measurement vector be $y=[y_1,y_2]^\intercal$. The state-space representation of the inverse kinematic problem is written as
\scriptsize
\begin{align*}
\textbf{x}_{i+1} &= \textbf{x}_i + \textbf{w}_i\\
\textbf{y}_i &= \begin{bmatrix}
\cos(\alpha_1)-\cos(\alpha_1+\alpha_2)\\[0.3em]
\sin(\alpha_1)-\sin(\alpha_1+\alpha_2)\\[0.3em]
\end{bmatrix}\begin{bmatrix}
r_1\\[0.3em]
r_2\\[0.3em]
\end{bmatrix} + \textbf{v}_k
\end{align*}
\normalsize
with zero-mean Gaussian process and measurement noises, $\textbf{w} \sim \mathcal{N}(\textbf{0},\rm{diag}[0.01^2,0.1^2])$ and $\textbf{w} \sim \mathcal{N}(\textbf{0},0.005\textbf{I}_2)$ respectively, where $\textbf{I}_{2} $ is the 2D identity matrix.

We compare the nonlinear filter performances of cubature Kalman filter and FBF using the root-mean square error (RMSE) of the angles over 200 Monte Carlo runs. As a self-assessment of its estimation errors, a filter provides an error covariance. Hence, we consider the filter-estimated RMSE as the square-root of the averaged appropriate diagonal entries of the covariance. The filter estimate is refereed to be consistent if the (true) RMSE is equal to its estimated RMSE. Again, we see that FBF outperforms CKF.

\begin{figure}[h]
	\centering
	\includegraphics[width=0.5\textwidth]{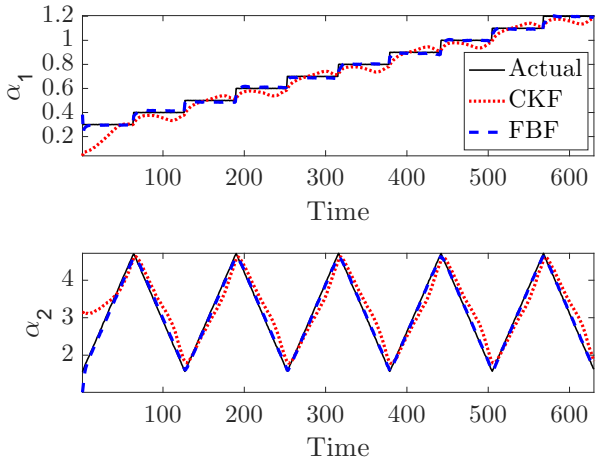}
	\caption{Inverse Kinematics State Estimation}
	\label{fig:arm_pred}
\end{figure} 
\begin{figure}[h]
	\centering
	\includegraphics[width=0.5\textwidth]{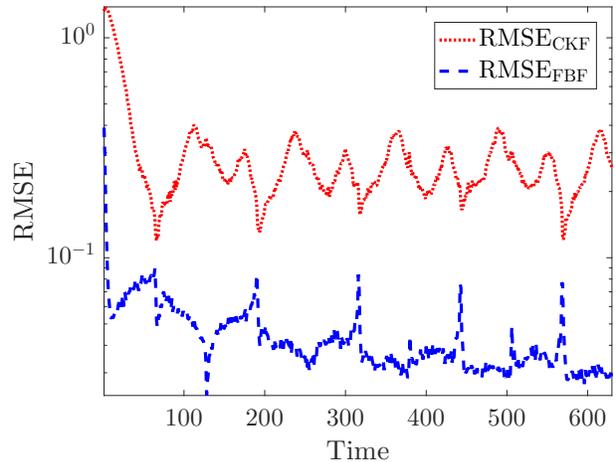}
	\caption{Performances}
	\label{fig:arm_RMSE}
\end{figure}

\section{Conclusion}
Various aspects of the classic Kalman filter have been extended for practical applications. This paper presents a novel formulation that tackles all three major shortcomings--linearity, prior knowledge of accurate system parameters, and Gaussian assumption--under a unifying framework, using the theory of reproducing kernel Hilbert Space. Applying kernel method and the representer theorem to perform linear quadratic estimation on a full state space model in a functional space, we derive the Bayesian recursive state estimation for a general nonlinear dynamical system in the original input space. Unlike existing nonlinear Kalman extensions where the system dynamics are assumed known, the state-space representation for the Functional Bayesian Filter is completely learned from observation, with universal approximation property. Using positive definite kernel function satisfying Mercer's conditions to compute information quantities, the FBF exploits both the statistical and time-domain information about the signal, extracts higher-order moments, and preserves the properties of covariances without the ill effects due to conventional arithmetic operations. This novel kernel adaptive filtering algorithm is applied to chaotic time-series estimation and prediction using Gaussian and non-Gaussian noises and inverse kinematics modeling. The simulation results show that it outperforms existing algorithms under different noise conditions.

Kernel method is extremely versatile and comes with many appealing properties. In the future, we will examine sparsification techniques for FBF, apply FBF to nonnumerical data such as graphs and modeling biological systems using neural spike trains, and explore applications for nonstationary environments using ideas from kernel transfer learning.
\bibliographystyle{plain}        
\bibliography{autosam}           
\end{document}